\documentclass[12pt]{article}
\usepackage{epsfig}
\usepackage{a4}
\usepackage{latexsym}

\newcommand\beq{\begin{equation}}
\newcommand\eeq{\end{equation}}
\newcommand\bea{\begin{eqnarray}}
\newcommand\eea{\end{eqnarray}}
\newcommand{\nn}{\nonumber}
\newcommand{\nin}{\noindent}

\newcommand\ra{\rightarrow}

\newcommand{\lsim}{\raisebox{-0.07cm}{$\:\stackrel{<}{{\scriptstyle
 \sim}}\: $} }

\newcommand{\als}{\alpha_{\rm s}}
\newcommand{\as}{a_{\rm s}}
\newcommand{\nf}{n^{}_{\! f}}
\newcommand{\nfs}{n^{\,2}_{\! f}}
\newcommand{\nft}{n^{\,3}_{\! f}}
\newcommand{\NmLO}{N$^{\rm m}$LO}
\newcommand{\LL}{L_{\Lambda}}

\newcommand{\MSbar}{$\overline{\mbox{MS}}$}

\newcommand\qV{\mbox{\boldmath $q$}}
\newcommand\eV{\mbox{\boldmath $e$}}
\newcommand\PV{\mbox{\boldmath $P$}}
\newcommand\RV{\mbox{\boldmath $R$}}
\newcommand\UV{\,\mbox{\boldmath $U$}}
\newcommand\IV{\,\mbox{\boldmath $I$}}
\newcommand\LV{\,\mbox{\boldmath $L$}}

\begin{document}
\setlength{\parskip}{0.3cm}
\setlength{\baselineskip}{0.55cm}

\begin{titlepage}

\noindent
NIKHEF 04-011 \hfill {\tt hep-ph/0408244} \\
August 2004 \\
\vspace{1.5cm}
\begin{center}
\Large
{\bf Efficient evolution of unpolarized and polarized} \\
\vspace{0.1cm}
{\bf parton distributions with QCD-PEGASUS} \\
\large
\vspace{3.2cm}
{\bf A. Vogt} \\
\vspace{1.2cm}
\normalsize
{\it NIKHEF Theory Group \\
\vspace{0.1cm}
Kruislaan 409, 1098 SJ Amsterdam, The Netherlands}
\vfill
\large
{\bf Abstract} \\
\end{center}
\vspace{-0.3cm}
\normalsize
The {\sc Fortran} package QCD-{\sc Pegasus} is presented. This program
provides fast, flexible and accurate solutions of the evolution 
equations for unpolarized and polarized parton distributions of hadrons 
in perturbative QCD. The evolution is performed using the symbolic
moment-space solutions on a one-fits-all Mellin inversion contour. 
User options include the order of the evolution including the 
next-to-next-to-leading order in the unpolarized case, the type of the 
evolution including an emulation of brute-force solutions, the evolution
with a fixed number $\nf$ of flavours or in the variable-$\nf$ scheme, 
and the evolution with a renormalization scale unequal to the 
factorization scale. 
The initial distributions are needed in a form facilitating the
computation of the complex Mellin moments.
\vspace*{0.5cm}

\end{titlepage}

\section*{Program Summary}

{\em Title of program\/}: QCD-{\sc Pegasus} \\[2mm]
{\em Version\/}: 1.0 \\[2mm]
{\em Catalogue identifier\/}: \\[2mm]
{\em Program obtainable from\/}: {\tt http://arxiv.org/archive/hep-ph} 
and its mirror sites by downloading the source of {\tt hep-ph/0408244}
\\[2mm]
{\em Distribution format\/}: uuencoded compressed tar file \\[2mm]
{\em E-mail\/}: {\tt avogt@nikhef.nl} \\[2mm]
{\em License\/}: GNU Public License \\[2mm]
{\em Computers\/}: all \\[2mm]
{\em Operating systems\/}: all \\[2mm]
{\em Program language\/}: {\sc Fortran 77} \\[2mm]
{\em Memory required to execute\/}: negligible ($<$ 1 MB) \\[2mm]
{\em Other programs called\/}: none \\[2mm]
{\em External files needed\/}: none \\[2mm]
{\em Number of bytes in distributed program, including test data 
     etc.\/}: 240$\,$578\\[2mm]
{\em Keywords\/}: unpolarized and longitudinally polarized parton 
distributions, Altarelli-Parisi evolution equations, Mellin-space 
solutions
\\[2mm]
{\em Nature of the physical problem\/}:
solution of the evolution equations for the unpolarized and polarized
parton distributions of hadrons at leading order (LO), next-to-leading 
order and next-to-next-to-leading order of perturbative QCD. Evolution 
performed either with a fixed number $\nf$ of effectively massless 
quark flavours or in the variable-$\nf$ scheme. The calculation of
observables from the parton distributions is not part of the
present~package.\\[2mm]
{\em Method of solution\/}:
analytic solution in Mellin space (beyond LO in general by power-%
expansion around the lowest-order expansion) followed by a fast Mellin
inversion to \mbox{$x$-space} using a fixed one-fits-all contour.
\\[2mm]
{\em Restrictions on complexity of the problem\/}: The initial 
distributions for the evolution are required in a form facilitating an
 efficient calculation of their complex Mellin moments. The ratio of 
the renormalization and factorization scales $\mu_r/\mu$ has to be a 
fixed number.
\\[2mm]
{\em Typical running time\/}: one to ten seconds, on a PC with a 2.0 
GHz Pentium-IV processor, for performing the evolution of 200 initial 
distributions to 500 $(x,\mu)$ points each. For more details see 
section \ref{sec-acc}.
\newpage

%
%
\section{Introduction}
%
%
Parton distributions form indispensable ingredients for analyses of 
hard processes with initial-state hadrons, investigated in fixed-target
experiments and at colliders like HERA, RHIC, TEVATRON and the 
forthcoming LHC. The task of determining these distributions can, in 
principle, be divided in two steps. The first is the determination of 
the non-perturbative initial distributions at some (usually rather low) 
scale $\mu_0$. The second is the perturbative calculation of their
scale dependence (evolution) to obtain the results at the hard 
scales $\mu$. For the foreseeable future, the non-perturbative input
cannot be calculated from first principles with  a sufficient accuracy.
Instead the initial distributions have to be fitted -- and re-fitted
once new data become available -- using a suitable set of 
hard-scattering observables. In practice the evolution thus enters in
both steps. 

Two methods for solving the evolution equations have been most widely 
applied in parton-distribution analyses. The first is the direct 
numerical integration of these integro-differential equations in
$(x,\mu)$-space, where $x$ stands for the momentum fraction carried by
the partons. Publicly available programs of this type can, for example,
be found in refs.~\cite{Miyama:1995bd,Hirai:1997gb} and \cite{QCDNUM}.
In the second approach a Mellin transformation is applied to turn the
evolution equations into systems of ordinary differential equations
(depending on the Mellin variable~$N$) which are more easily accessible 
to a further analytic treatment. A {\tt C++} code of this type has been 
published in ref.~\cite{Weinzierl:2002mv}. More programs of both types 
have been used and/or informally circulated in the perturbative-QCD 
community.

In this article we present another $N$-space evolution package. For 
easier reference the program has been given a name, {\sc QCD-Pegasus} 
or {\sc Pegasus} in short, standing for `Parton Evolution Generated 
Applying Symbolic $U$-matrix Solutions'. Here \mbox{`$U$-matrix'} is a 
usual name for the key technical ingredient in the solutions which 
dates back, at least, to ref.~\cite{FP82}.
The present program is a descendant of the code written by the
present author fifteen years ago for the GRV analyses started in 
ref.~\cite{Gluck:1989ze}. Since then it has been almost completely 
rewritten more than once. Various intermediate versions have been 
regularly used in QCD studies by the author and by others. The present
last incarnation of the program includes, in a now hopefully 
sufficiently well documented and user-friendly manner, the evolution of 
unpolarized and helicity-dependent parton densities up to (in the former case) 
the next-to-next-to-leading order (N$^2$LO) of perturbative QCD for any 
fixed ratio of the factorization and renormalization scales. 
The user can choose between various ways to truncate contributions of 
higher order (including emulations of the brute-force solutions) in 
both the fixed flavour-number and the variable-$\nf$ evolution schemes.

This manual is organized as follows. In section 2 we recall the 
formalism used for the $N$-space evolution of the parton distributions.
Topics related to the inverse Mellin transformations of the solutions
and the initial distributions are then discussed in section~3. A compact 
user guide for the program is provided in section 4, followed in 
section 5 by a short reference manual of all routines. In section 6 we
briefly address the accuracy and speed of the evolution by 
{\sc QCD-Pegasus}, before we conclude in section 7. 
%
%
 
\setlength{\baselineskip}{0.53cm}
\section{The evolution equations and their solution}
In this section we discuss, in some detail, the formalism employed in 
the program for the Mellin-space solution of the evolution equations. 
In the course of the discussion we point out some default choices made 
in the present version of the program, explain the main options 
available to the user, and indicate some structural restrictions.
%
%
\subsection{The running coupling constant}
\label{sec-as}
%
%
As the reader will see below, the strong coupling constant $\as$ plays
a more central role in the present approach to the evolution of parton
densities than usually in $x$-space programs. We therefore start the 
discussion with 
$\as$ for which we employ the normalization
\beq
\label{as-def}
  \as \:\equiv\: \frac{\als}{4\pi} \:\: .
\eeq
At \NmLO\ the scale dependence of $\as$ is given by
\beq
\label{as-eqn}
  \frac{d\, \as}{d \ln \mu_r^2} \: = \: \beta_{\:\!\rm N^mLO}(\as)
  \: = \: - \sum_{k=0}^m \, \as^{k+2} \,\beta_k \:\: ,
\eeq
where $\mu_r$ denotes the renormalization scale and $\nf$ stands for the
number of effectively massless quark flavours. $\nf$ is considered a 
fixed number until section \ref{sec-hflav}. The expansion coefficients
$\beta_k$ of the $\beta$-function of QCD are known up to $k=3$, i.e.,  
N$^3$LO
\bea
\label{beta-exp}
  \beta_0 &\, =\, & \: 11\: - \: 2/3\: \nf
  \nonumber \\
  \beta_1 &\, =\, & 102 - 38/3\, \nf
  \nonumber \\
  \beta_2 &\, =\, & \, 2857/2\, - 5033/18\, \nf + \,325/54\: \nfs
  \nonumber \\
  \beta_3 &\, =\, & 29243.0 - \: 6946.30\: \nf + 405.089\, \nfs
                  + 1093/729\, \nft \:\: .
\eea
Here the scheme-dependent quantities $\beta_2$ \cite{Tarasov:1980au,%
Larin:1993tp} and $\beta_3$ \cite{vanRitbergen:1997va} refer to the 
usual \MSbar\ scheme. Only this scheme has been implemented in the 
program so far. For brevity the irrational coefficients of $\beta_3$ in 
Eq.~(\ref{beta-exp}) have been truncated to a sufficient accuracy of 
six digits.

Eq.~(\ref{as-eqn}) can be integrated in a closed form only at low 
orders, and even then one only arrives at an implicit equation for 
$\as(\mu_r^2)$ beyond LO. The exact solution at NLO which expresses 
$\as(\mu_r^2)$ in terms of its value $\as(\mu_0^2)$ at a reference 
scale $\mu_0^2$, for example, reads
\beq
\label{as-impl}
  \frac{1}{\as(\mu_r^2)} \: = \: \frac{1}{\as(\mu_0^2)}  
  + \beta_0 \ln \left( \frac{\mu_r^2}{\mu_0^2} \right)
  - b_1 \ln \left\{ \frac{\as(\mu_r^2) \, [ 1 + b_1 \as(\mu_0^2) ]}
    {\as(\mu_0^2) \, [ 1 + b_1 \as(\mu_r^2) ]} \right\} 
\eeq
with $b_k\equiv \beta_k /\beta_0$. The program uses Eq.~(\ref{as-impl})
only at LO ($b_1 = 0$), as in general a numerical iteration is required
otherwise anyway. Beyond LO the value of $\as(\mu_r^2)$ is by default
determined directly from Eq.~(\ref{as-eqn}) by a fourth order 
Runge-Kutta integration~\cite{AbrSteg1}. 
 
All known orders $m \leq 3$ in Eq.~(\ref{as-eqn}) are available in the 
routine for $\as$ to which $m$ is transferred as {\tt NAORD}. In the 
context of the evolution program, {\tt NAORD} is not set externally but 
specified via the order chosen for the splitting functions, see section 
\ref{sec-pevol}.
 
Another very common approach, used for instance by the Particle Data 
Group \cite{Eidelman:2004wy}, is to expand the solution in inverse
powers of $\LL \equiv \ln (\mu_r^2/\Lambda^2)$ where $\Lambda$ 
is the QCD scale parameter. Up to N$^3$LO this expansion yields
\cite{Chetyrkin:1997sg} 
\bea
\label{as-exp1}
 \as(\mu_r^2) 
 &\!=\!& \frac{1}{\beta_0\LL} \: - \: 
    \frac{1}{(\beta_0\LL)^2} \, b_1\ln \LL \: +\: 
    \frac{1}{(\beta_0\LL)^3} \left[ \, b_1^2 \left( \ln^2 \LL-\ln \LL-1 
    \right) + b_2 \right] \nn\\
 &\!+\!& \frac{1}{(\beta_0\LL)^4}\left[ \, b_1^3 \left(-\ln^3 \LL + 
    \frac{5}{2} \ln^2\LL +2 \ln\LL-\frac{1}{2}\right) - 3b_1 b_2\ln\LL
    +\frac{b_3}{2} \,\right] \:\: . \qquad
\eea
Eq.~(\ref{as-exp1}) solves the evolution equation (\ref{as-eqn}) only 
up to higher orders in $1/\LL$. As explained in section \ref{sec-nevol},
this is an unwanted feature for the $N$-space evolution which especially
bedevils direct comparisons to $x$-space evolution programs. Therefore 
the use of Eq.~(\ref{as-exp1}) is not a standard option in the present
evolution~package.
%
%
\subsection{The general evolution equations}
\label{sec-pevol}
%
%
The scale dependence of the parton distributions is governed by the 
evolution equations 
\beq
\label{f-evol}
  \frac{\partial}{\partial \ln \mu^2}\, f_i(x,\mu^2) \: = \:
  P_{ij}(x,\mu^2) \otimes f_j(x,\mu^2) \:\: .
\eeq
Here $\mu$ represents the factorization scale, and for the moment we 
put $\mu_r = \mu$. $f_i(x,\mu^2)$ stands for the number distributions
of quarks, antiquarks and gluons in a hadron, where $x$ represents 
the fraction of the hadron's momentum carried by the parton. Summation
over the parton species $j$ is understood, and $\otimes$ stands for the 
Mellin convolution. Eq.~(\ref{f-evol}) thus represents a system of 
$2\nf\!+\!1$ coupled integro-differential equations. The \NmLO\ 
approximation for the splitting functions $P_{ij}(x,\mu^2)$ reads
\beq
\label{P-exp}
 P_{ij}^{\,\rm N^{\rm m}LO}(x,\mu^2) \: =\:
     \sum_{k=0}^{m} \, \as^{k+1} (\mu^2) \, P_{ij}^{(k)}(x) \:\: .
\eeq
The splitting functions for the spin-averaged (unpolarized) case are 
now known at N$^2$LO ($\equiv$ NNLO) \cite{Moch:2004pa,Vogt:2004mw}.
Note that $P_{ij}(x,\mu^2)$ depends on $\mu$ only via the coupling
$\as(\mu^2)$, a feature which forms the basis for the $N$-space 
solution of Eq.~(\ref{f-evol}) discussed in section~\ref{sec-nevol}.

The splitting functions for the general case $\mu_r \neq \mu$ can be
obtained from Eq.~(\ref{P-exp}) by Taylor-expanding $\as(\mu^2)$ in
terms of $\as(\mu_r^2)$. Up to N$^3$LO this leads to
\bea
\label{P-exp2}
  P_{ij}(x,\mu,\mu_r) 
  & \!\! = \!\! & 
      \as(\mu_r^2) \: P_{ij}^{(0)}(x) \nn \\[0.5mm]
  & + & 
    \as^2(\mu_r^2) \, \Big( P_{ij}^{(1)}(x) - \beta_0
      P_{\rm NS}^{(0)}(x) L \Big) \: \\[0.5mm]
  & + & 
    \as^3(\mu_r^2) \, \Big( P_{ij}^{(2)}(x)
    - 2\beta_0 L \, P_{ij}^{(1)}(x) 
    - \left\{ \beta_1 L - \beta_0^2 L^2 \right\} P_{ij}^{(0)}(x) 
    \Big) \nn\\[0.5mm] 
  & + & 
    \as^4(\mu_r^2) \, \Big( P_{ij}^{(3)}(x)
    - 3\beta_0 L \, P_{ij}^{(2)}(x) 
    - \left\{ 2\beta_1 L - 3\beta_0^2 L^2 \right\} P_{ij}^{(1)}(x) 
     \nn \\
  &   & \qquad\qquad
    - \left\{ \beta_2 L - 5/2\, \beta_1 \beta_0 L^2 + \beta_0^3 L^3
    \right\} P_{ij}^{(0)}(x) \Big) \nn
\eea
with $L \equiv \ln (\mu^2 / \mu_r^2)$. If $L$ is a fixed number, then 
also in Eq.~(\ref{P-exp2}) the coefficients of $\as^k(\mu_r^2)$ depend 
only on $x$, and the algorithms described below are applicable. In 
other words, the program is not designed to deal with choices like
$\mu_r^2 = M^2 + \mu^2$ where $M$ is some mass scale. Note, however,
that no such restriction is in place between physical scales and $\mu$.

The order $m$ in Eq.~(\ref{P-exp}) (denoted by {\tt NPORD}) and the
ratio $\mu^2 / \mu_r^2$ (denoted by {\tt FR2}) are initialization
parameters of the evolution package. The values $m = 0, 1$ and 2 are 
available at present for the standard \MSbar\ factorization scheme. An 
extension to $m=3$ --- based on future partial results or even Pad{\'e} 
estimates for $P_{ij}^{(3)}$ --- may be useful for uncertainty 
estimates in special cases, e.g., in determinations of $\as$ from 
structure functions~\cite{vanNeerven:2001pe}.
%
%
\subsection{The flavour decomposition}
\label{sec-lflav}
%
%
It is convenient to decompose the system (\ref{f-evol}) as far as 
possible from charge conjugation and flavour symmetry constraints alone.
The gluon-quark and quark-gluon splitting functions are flavour
independent
\beq
\label{Poffd}
  P_{\rm gq} \: \equiv \: P_{{\rm gq}_{i}} 
             \: = \: P_{{\rm g}\bar{\rm q}_{i}} \:\: , \quad
  P_{\rm qg} \: \equiv \: \nf\, P_{{\rm q}_{i}\rm g} 
             \: = \: \nf\, P_{\bar{\rm q}_{i}\rm g} \:\: .
\eeq
Any difference $q_i - q_j$ and $q_i - \bar{q}_j$ of quark and 
(anti-)quark distributions therefore decouples from the gluon density 
$g$. Hence the combination maximally coupling to $g$ is the 
flavour-singlet quark distribution
\beq
\label{q-sg}
  q_{\:\!\rm s} \: = \: \sum_{r=1}^{\nf} \, (q_r + \bar{q}_r) 
\eeq
evolving according to 
\beq
\label{ev-sg}
  \frac{d}{d \ln\mu^2}
  \left( \begin{array}{c} \! q_{\:\!\rm s} \! \\ g  \end{array} 
  \right) \: = \: \left( \begin{array}{cc} P_{\rm qq} & P_{\rm qg} \\
  P_{\rm gq} & P_{\rm gg} \end{array} \right) \otimes \left( 
  \begin{array}{c} \!q_{\:\!\rm s}\! \\ g  \end{array} \right)
  \:\: .
\eeq
The singlet quark-quark splitting function $P_{\rm qq}$ is specified in
Eq.~(\ref{Pqq}) below.

In order decouple the non-singlet (difference) combinations, we make 
use of the general structure of the (anti-)quark (anti-)quark splitting 
functions,
\bea
\label{p-symm}
  P_{{\rm q}_{i}{\rm q}_{k}} \: = \:
  P_{\bar{{\rm q}}_{i}\bar{{\rm q}}_{k}}
  &\! =\! & \delta_{ik} P_{{\rm q}{\rm q}}^{\,\rm v}
        + P_{{\rm q}{\rm q}}^{\,\rm s} \nonumber \\
  P_{{\rm q}_{i}\bar{{\rm q}}_{k}} \: = \:
  P_{\bar{{\rm q}}_{i}{\rm q}_{k}}
  &\! =\! & \delta_{ik} P_{{\rm q}\bar{{\rm q}}}^{\,\rm v}
        + P_{{\rm q}\bar{{\rm q}}}^{\,\rm s}
  \:\: .
\eea
In general (beyond NLO), Eq.~(\ref{p-symm}) leads to three independently
evolving types of non-singlet combinations. The flavour asymmetries 
$q_{\rm ns}^{\,\pm}$ and the total valence distribution 
$q_{\rm ns}^{\rm v}$,
\beq
\label{q-ns}
  q_{{\rm ns},ik}^{\,\pm} \: = \: q_i \pm \bar{q}_i 
  - (q_k \pm \bar{q}_k) \:\: , \quad q_{\rm ns}^{\rm v} \: = \: 
  \sum_{r=1}^{\nf} \, (q_r - \bar{q}_r) \:\: ,
\eeq
respectively evolve with
\bea
\label{P-ns}
  P_{\rm ns}^{\,\pm} & = & P_{{\rm q}{\rm q}}^{\,\rm v}
    \pm P_{{\rm q}\bar{{\rm q}}}^{\,\rm v} \:\: , \nn \\
  P_{\rm ns}^{\,\rm v} & = & P_{\rm qq}^{\,\rm v}
    - P_{{\rm q}\bar{{\rm q}}}^{\,\rm v} + \nf (P_{\rm qq}^{\,\rm s}
    - P_{{\rm q}\bar{{\rm q}}}^{\,\rm s}) \: \equiv \:
    P_{\rm ns}^{\, -} + P_{\rm ns}^{\,\rm s} \:\: .
\eea
Finally the singlet splitting function (\ref{Pqq}) can be expressed as
\beq
\label{Pqq}
  P_{\rm qq} \: =\: P_{\rm ns}^{\,+} + \nf (P_{\rm qq}^{\:\rm s}
  + P_{\rm {\bar{q}q}}^{\:\rm s})
  \:\equiv\:  P_{\rm ns}^{\,+} + P_{\rm ps}^{} \:\: .
\eeq
In the  expansion in powers of $a_{\rm s}$, the flavour-diagonal 
(`valence') quantity $P_{\rm qq}^{\,\rm v}$ in Eq.~(\ref{p-symm}) 
starts at first order. $P_{{\rm q}\bar{\rm q}}^{\,\rm v}$ and the 
flavour-independent (`sea') contributions $P_{\rm qq}^{\,\rm s}$ and 
$P_{{\rm q}\bar{{\rm q}}}^{\,\rm s}$ --- and hence the `pure-singlet' 
term $P_{\rm ps}$ in Eq.~(\ref{Pqq}) --- are of order $\as^2$.
A non-vanishing $P_{\rm ns}^{\,\rm s} \sim P_{{\rm qq}}^{\,\rm s} - 
P_{{\rm q} \bar{{\rm q}}}^{\,\rm s}$ in Eq.~(\ref{P-ns}) occurs for 
the first time at the third order.

For the evolution of the flavour asymmetries in Eq.~(\ref{q-ns}) we use
the basis
\beq
\label{ns-basis}
  v_l^{\pm} \: = \: \sum_{i=1}^{k} (q_{i} \pm \bar{q}_{i}) - k (q_{k} 
  \pm \bar{q}_{k}) 
\eeq
with $ k = 1,\ldots ,\nf $ and the usual group-theoretical notation 
$l = k^{2} - 1$. After performing the evolution, the individual quark 
and antiquark distributions can be recovered using
\beq
   q_{i} + \bar{q}_{i} \: = \: 
   \frac{1}{\nf}\, q_{\:\!\rm s} - \frac{1}{i}\, v_{i^{2}-1}^+
   + \sum_{k=i+1}^{\nf} \frac{1}{k(k-1)}\, v_{k^{2}-1}^+ 
\eeq
where $ v_0^+ \equiv 0 $, together with the corresponding equation for 
the differences $q_{i} - \bar{q}_{i}$.
%
%
\subsection{The \boldmath{$N$}-space solutions}
\label{sec-nevol}
%
%
In the next two sections we describe the algorithm employed for the 
solution of the evolution equations in Mellin-$N$ space. Thus we now 
switch to the moments of all \mbox{$x$-dependent} quantities,
\beq
\label{N-trf}
   a(N) \: = \: \int_0^1 \! dx \: x^{N-1}\, a(x) \:\: .
\eeq
The advantage of this transformation is that is turns the Mellin
convolutions into simple products,
\beq
\label{N-conv}
   [ a \otimes b ](N) \: = \: a(N)\, b(N) \:\: ,
\eeq
which greatly simplifies all further manipulations. The disadvantage of
working in \mbox{$N$-space} is that all quantities have to be known for
complex values of $N$ for the final transformation back to $x$-space.
The resulting limitations of the program are discussed in section 3.

As discussed above, we restrict ourselves to situations where the scale 
$\mu$ enters the right-hand side of Eq.~(\ref{P-exp2}) only through the 
(monotonous) coupling $\as \equiv \as(\mu_r^2 \! =\! \kappa \mu^2)$.
Hence we can switch to $\as$ as the independent variable. Using a 
matrix notation for the singlet system (\ref{ev-sg}), the combination
of Eqs.~(\ref{as-eqn}) and (\ref{P-exp}) yields 
\bea
\label{evol1}
 \frac{\partial \qV (N,\as)}{\partial \as}
 &\!=\!& \left\{ \beta_{\:\!\rm N^mLO}(\as) \right\} ^{-1} 
     \PV_{\rm N^{\rm m}LO}(N,\as) \: \qV (N,\as) \nn\\
 &\!=\!& -\frac{1}{\beta_0 \as} \left[ \PV^{(0)}(N) + \as 
     \Big( \PV^{(1)}(N)- b_1 \PV^{(0)}(N)\Big) \right.  \nn \\ 
 &     & \qquad\quad \left. \mbox{}
     + \as^2 \left( \PV^{(2)}(N) - b_1 \PV^{(1)}(N) + ( b_1^2-b_2 ) 
     \PV^{(0)}(N) \right) + \ldots \right] \, \qV (N,\as) \quad \nn \\
 &\!=\!& -\frac{1}{\as} \bigg[ \RV_0 (N) + \sum_{k=1}^{\infty} \as^k 
     \RV_k (N) \bigg] \, \qV (N,\as) \:\: .
\eea
In the last line we have introduced the recursive abbreviations
\beq
  \RV_0 \:\equiv\: \frac{1}{\beta_0} \PV^{(0)} \:\: , \quad
  \RV_k \:\equiv\: \frac{1}{\beta_0} \PV^{(k)} - \sum_{i=1}^{k}
                    b_i \,\RV_{k-i}
\label{rmat}
\eeq
for the splitting function combinations entering the new expansion
(\ref{evol1}). $b_k$ has been defined below Eq.~(\ref{as-impl}), and
$\PV^{(k)}$ is the coefficient of $\as^{\:\!k+1}$ in Eq.~(\ref{P-exp2}).
As in Eq.~(\ref{rmat}), we will often suppress the explicit reference 
to the Mellin variable $N$ below.

The singlet splitting-function matrices $\RV_k $ of different orders 
$k$ do not commute. Therefore the solution of Eq.~(\ref{evol1}) cannot
be written in a closed exponential form beyond LO. Instead we employ
a series expansion around the lowest order solution,
\beq
\label{lo-sol}
 \qV_{\,\rm LO} (N,\as,N) \: = \:
 \left( \frac{\as}{a_0} \right)^{{\footnotesize -\RV}_0 (N)}
 \qV (N,a_0) \equiv \LV (N,\as,a_0) \, \qV (N,a_0) 
\eeq
with $a_0 \equiv \as(\mu_{r,0}^2 \! =\! \kappa \mu_0^2)$. This 
expansion reads
\bea
\label{u-sol}
 \qV (N,\as) &=& \UV (N,\as) \LV (N,\as,a_0) \UV^{-1}(N,a_0) \:
   \qV (N,a_0) \\[0.5mm]
  &=& \bigg[\, 1 + \sum_{k=1}^{\infty} \as^k \UV_{\!k} (N) \bigg] \LV 
    (\as,a_0,N) \bigg[\, 1 + \sum_{k=1}^{\infty} a_0^k \UV_{\!k} (N) 
    \bigg]^{-1} \qV (a_0,N) \:\: . \nn
\eea
Here the third, $a_s$-independent factor normalises the evolution 
operator to the unit matrix at $\mu_0^2$, instead of to the LO result 
(\ref{lo-sol}) at infinitely high scales. The evolution matrices 
$\UV_{\!k}$ are constructed from the splitting function combinations
$\RV_{\,l\leq k}$ in the next section.

Eqs.~(\ref{evol1}) and (\ref{u-sol}) offer various ways to define the
\NmLO\ solution which differ in terms of order $n>m$. In the program
the choice between the resulting options is made by the initialization
parameter {\tt IMODEV}. 
One obvious choice is to keep the terms originating from $\beta_
{\:\!\rm N^mLO}$ in Eq.~(\ref{as-eqn}) and $\PV_{\rm N^{\rm m}LO}$ in
Eq.~(\ref{P-exp}) to all orders (in practice: to a sufficiently high
order) in both (\ref{evol1}) and (\ref{u-sol}). This is equivalent to
a direct iterative solution of Eq.~(\ref{f-evol}) as performed by 
standard $x$-space evolution programs, to which the results can then
be compared directly. This mode for the evolution is invoked for
{\tt IMODEV} = 1.

Note that this equivalence only holds if $\as(\mu_r^2)$ solves Eq.\
(\ref{as-eqn}) exactly. Otherwise, for example if Eq.~(\ref{as-exp1}) 
is used, the evolution equation (\ref{evol1}) in $\as$ is obtained from 
Eq.~(\ref{f-evol}) by dividing the l.h.s.\ and the r.h.s.\ by 
quantities which differ somewhat, viz $d \as / d \ln \mu_r^2$ and 
$\beta_{\:\!\rm N^mLO}$. 
The difference introduced by this mismatch is a higher-order effect, 
but its numerical impact is far from negligible, e.g., most of the 
difference between the two parts of table~1 in the 1996 NLO comparisons
\cite{Blumlein:1996rp} arises from this source.  

One can equally well take the point of view that at \NmLO\ terms beyond 
$\as^m$ should be removed in the square brackets in Eq.~(\ref{evol1}),
as these terms would receive contributions from $\PV^{(n>m)}$ and 
$\beta_{n>m}$. At N$^2$LO, for example, one then retains only the terms
explicitly written down in the second and third line of 
Eq.~(\ref{evol1}). If still `all' orders are kept in the solution
(\ref{u-sol}), one arrives at a second iterative option employed for
{\tt IMODEV} = 2. Finally one can instead apply the same reasoning to 
the matrices $\UV_{n>m}$ in Eq.~(\ref{u-sol}) which would also receive
contributions from $\PV^{(n>m)}$ and $\beta_{n>m}$. Expanding also the
$\UV^{-1}$ term in Eq.~(\ref{u-sol}), necessary in the singlet case 
as explained below Eq.~(\ref{U-res}), one then arrives at the so-called 
truncated solution. Up to N$^3$LO this solution is thus given by 
\cite{Ellis:1993rb}
\bea
\label{f-trnc}
 \qV_{\rm N^3LO} (\as) 
 &=& \Big[ \LV + \as \UV_{\!1} \LV - a_0 \LV \UV_{\!1} \nn \\
 & & \mbox{} + \as^2 \UV_{\!2} \LV - \as a_0 \UV_{\!1} \LV \UV_{\!1} 
     + a_0^2 \LV \left( \UV_{\!1}^2 - \UV_{\!2} \right) \nn \\
 & & \mbox{} + \as^3 \UV_{\!3} \LV - \as^2 a_0 \UV_{\!2} \LV \UV_{\!1}
     + \as a_0^2 \UV_{\!1} \LV \left( \UV_{\!1}^2 - \UV_{\!2} \right)
     \nn \\
 & & \quad \mbox{} - a_0^3 \LV \left( \UV_{\!1}^3 - \UV_{\!1} \UV_{\!2} 
     - \UV_{\!1} \UV_{\!2}+ \UV_{\!3} \right) \Big] \: \qV (a_0) \:\: , 
\eea
where we have suppressed all arguments of $\LV(N,\as,a_0)$ for brevity.
The NLO (NNLO) approximations are obtained from Eq.~(\ref{f-trnc}) by
respectively retaining only the first (first and second) line in the 
square bracket. These truncated \NmLO~solutions (at present implemented
for $m \leq 2$) are employed by the program for any value of 
{\tt IMODEV} other than 1, 2 and, for the non-singlet cases, 3. The
latter case is addressed in section \ref{sec-nsing}.
 
The approaches discussed above obviously differ only in terms beyond 
the order under consideration. The iterative procedures introduce more 
scheme-dependent higher-order terms into the evolution of observables 
in a general factorization scheme. 
On the other hand, the truncated solution does not satisfy the 
evolution equations (\ref{f-evol}) literally, but only in the sense of 
a power expansion, i.e., up to terms of order $n>m$ like 
Eq.~(\ref{as-exp1}) for the coupling constant. 
The differences between the respective results can be regarded as a 
lower limit for the uncertainties due to higher-order terms. 
%
%
\subsection{The evolution matrices \boldmath{$\UV_{\!k}$}}
\label{sec-umat}
%
%
Inserting the ansatz (\ref{u-sol}) into the evolution equations (\ref
{evol1}) and sorting in powers of $\as$, one arrives at a chain of 
commutation relations for the expansion coefficients $\UV_{\!k}(N)\,$:
\bea
\label{U-eqn}
  \big[ \UV_{\!1}, \RV_0 \big] & = & \RV_1 + \UV_{\!1}  \nn \\
  \big[ \UV_{\!2}, \RV_0 \big] & = & 
     \RV_2 + \RV_1 \UV_{\!1} + 2 \UV_{\!2}  \\
  & \vdots & \nn \\
  \big[ \UV_{\!k}, \RV_0 \big] & = & 
     \RV_k + \sum_{i=1}^{k-1} \RV_{k-i} \UV_i + k \UV_{\!k}
  \: \equiv \: \widetilde{\RV}_k + k \UV_{\!k} \:\: . \nn
\eea
For the flavour-singlet system (\ref{ev-sg}) these equations can be 
solved recursively by applying the eigenvalue decomposition of the LO 
splitting function matrix~\cite{Ellis:1993rb,Blumlein:1997em}, 
completely analogous to the classical truncated NLO solution with only 
$\UV_1$ in ref.~\cite{FP82}. 
One writes
\beq
\label{lo-dec1}
  \RV_0 = r_{-} \eV_{-} + r_{+} \eV_{+} \:\: ,
\eeq
where $r_{-}$ ($r_{+}$) stands for the smaller (larger) eigenvalue of
$ \RV_0 $,
\beq
\label{lo-dec2}
  r_{\pm } = \frac{1}{2 \beta_0} \bigg[ P_{qq}^{(0)} + P_{gg}^{(0)}
  \pm \sqrt{ \Big( P_{qq}^{(0)} - P_{gg}^{(0)}\Big) ^2 +
  4 P_{qg}^{(0)} P_{gq}^{(0)} } \, \bigg] \:\: .
\eeq
The matrices $\eV_{\pm }$ denote the corresponding projectors,
\beq
\label{lo-dec3}
 \eV_{\pm} = \frac{1}{r_{\pm}-r_{\mp}} \Big[ \RV_0-r_{\mp}\IV \Big]
 \:\: ,
\eeq
and $\IV $ the $2\!\times\! 2 $ unit matrix. Thus the LO evolution 
operator (\ref{lo-sol}) can be represented~as
\beq
\label{lo-exp}
 \LV (a_s,a_0,N) = \eV_{-}(N) \bigg(\frac{a_s}{a_0}\bigg)^{-r_{-}(N)}
   + \eV_{+}(N) \bigg(\frac{a_s}{a_0}\bigg)^{-r_{+}(N)} \: .
\eeq
Inserting the identity
\beq
  \UV_{\!k} = \eV_{-} \UV_{\!k} \eV_{-} + \eV_{-} \UV_{\!k} \eV_{+} 
            + \eV_{+} \UV_{\!k} \eV_{-} + \eV_{+} \UV_{\!k} \eV_{+}
\eeq
into the commutation relations (\ref{U-eqn}), one finally obtains the
coefficients in Eq.\ (\ref{u-sol}),
\beq
\label{U-res}
  \UV_{\!k} =
  - \frac{1}{k} \Big[ \eV_{-} \widetilde{\RV}_k \eV_{-} +
    \eV_{+} \widetilde{\RV}_k \eV_{+} \Big]
  + \frac{\eV_{+} \widetilde{\RV}_k \eV_{-}}{r_{-} - r_{+} - k}
  + \frac{\eV_{-} \widetilde{\RV}_k \eV_{+}}{r_{+} - r_{-} - k} \: .
\eeq
Note that the poles in $\UV_k (N)$ at $N$-values where $r_{-}(N) - 
r_{+}(N) \pm k $ vanishes are cancelled by the $\UV^{-1} $ term in the 
solution (\ref{u-sol}). The expansion of $\UV^{-1} $ mentioned above
Eq.~(\ref{f-trnc}) achieves this cancellation also for the truncated 
solutions. This can be made directly visible by inserting the explicit
forms (\ref{lo-exp}) and (\ref{U-res}) into the solution (\ref{f-trnc})
and writing the contributions in an appropriate order. At NLO, for 
example, one arrives at
\bea
\label{nlo-exp}
  \qV(\as) &=& \Bigg\{ \left( \frac{\as}{a_0} \right)^{-r_-} 
    \bigg[\, \eV_- + (a_0 - \as)\:  \eV_- \RV_1 \eV_- \\ 
  & &  \qquad \mbox{}
    - \bigg( a_0 - \as \left( \frac{\as}{a_0} \right)^{r_--r_+} \bigg)
    \,\frac{\eV_- \RV_k \eV_+}{r_+-r_--1} \,\bigg] 
    \: + \: (\, + \leftrightarrow - \,) \Bigg\} \: \qV(a_0) \:\: . \nn
\eea
Not only the denominator in the second line vanishes for $r_+-r_-=1$,
but also its coupling-constant prefactor in the round brackets. 
%
%
\subsection{Non-singlet cases and symmetry breaking}
\label{sec-nsing}
%
%
Eq.~(\ref{U-eqn}) also holds for the scalar evolution of the non-%
singlet combinations (\ref{q-ns}) of the quarks distributions, but with 
the obvious simplification that the right-hand sides vanish. This
facilitates a direct recursive solution for $U_k^{\,\rm ns}$ in which, 
unlike the singlet results (\ref{U-res}), no spurious poles occur.
Consequently the truncated solution can be written down also without
the expansion of $\,U^{-1}$ in this case, at \NmLO\ yielding
\beq
\label{ns-trnc}
 q_{\rm ns}^{\,\pm,\rm v} (\as) \: = \:
  \bigg[\, 1 + \sum_{k=1}^m \as^k\, U^{\,\pm,\rm v}_{k} \bigg] 
  \bigg[\, 1 + \sum_{k=1}^m a_0^k\, U^{\,\pm,\rm v}_{k} \bigg]^{-1}
  \left( \frac{\as}{a_0} \right)^{-R_0^{\:\!\rm ns}}
  q_{\rm ns}^{\,\pm,\rm v} (a_0) \:\: . 
\eeq
This solution is accessed by {\tt IMODEV} = 3 (together with Eq.\
(\ref{f-trnc}) for the singlet case).

Both iterated non-singlet solutions can be written down in a compact 
closed form at NLO. Hence instead of the expansion (\ref{u-sol}) we 
use for {\tt IMODEV} = 1
\beq
\label{ns-sol1}
  q_{\rm NLO}^{\,\pm} (\as) \: = \: \exp \left\{ \,\frac{U^{\,\pm}_{1}}
  {b_1} \ln \left( \frac{1+b_1 \as}{1+b_1 a_0} \right) \right\} 
  \left( \frac{\as}{a_0} \right)^{-R_0^{\:\!\rm ns}} 
  q_{\rm ns}^{\,\pm}(a_0) \:\: ,
\eeq
and for {\tt IMODEV} = 2
\beq 
\label{ns-sol2}
  q_{\rm NLO}^{\,\pm} (\as) \: = \: \exp \left\{ (\as - a_0) \,
  U^{\,\pm}_{1} \right\} \left( \frac{\as}{a_0} \right)
  ^{-R_0^{\,\rm ns}} q_{\rm ns}^{\,\pm}(a_0) \:\: .
\eeq
At NNLO the non-singlet solutions have been programmed analogous to the
singlet case.

Due to the differences of $P_{\rm ns}^{\,+}$ and $P_{\rm ns}^{\,-}$ for 
$m \geq 1$ and of $P_{\rm ns}^{\,-}$ and $P_{\rm ns}^{\,\rm v}$ for 
$m \geq 2$ in Eq.~(\ref{P-ns}), qualitatively new effects arise at the 
respective orders $m$, namely a breaking of symmetries imposed on the 
initial distributions. 

Consider the NLO evolution of an input $u = u_{\rm v} + \bar{u}$, $d = 
d_{\rm v} + \bar{d}$ with an SU(2)-symmetric sea, $\bar{u}(\mu_0^2) = 
\bar{d}(\mu_0^2) $. The initial distributions for the evolution of 
$v_3^-$ and $v_3^+$ in Eq.~(\ref{ns-basis}) are then identical, but 
due to $R_1^+ \neq R_1^-$ not the results of the evolution,  
\bea
\label{ns-sol3}
  (u_{\rm v}-d_{\rm v})(\as) &=& \left\{\, 1 + ( \as - a_0 ) \, R_1^-
       \right\} \left( \frac{\as}{a_0} \right)^{-R_0^{\:\!\rm ns}}
       (u_{\rm v}-d_{\rm v})(a_0) \nn \\
  v_3^+(\as) &=&  \left\{\, 1 + ( \as - a_0 ) \, R_1^+
       \right\} \left( \frac{\as}{a_0} \right)^{-R_0^{\:\!\rm ns}}
       (u_{\rm v}-d_{\rm v})(a_0) \:\: .              
\eea
Subtracting these two (truncated) solutions yields
\beq
  2(\bar{u}-\bar{d})(\as) \: = \: ( \as - a_0 ) \, \left( R_1^+ - R_1^- 
  \right) \left( \frac{\as}{a_0} \right)^{-R_0^{\:\!\rm ns}}
  (u_{\rm v}-d_{\rm v})(a_0) \:\: .              
\eeq
Hence a flavour symmetry of the input sea quark densities is not 
preserved by the NLO evolution, if the valence distributions (as in the
case of the proton) are not identical. The same procedure applied to
$v_8^-$ and $q_{\rm v}$ in Eq.~(\ref{q-ns}) shows that $s \neq \bar{s}$ 
at NNLO even for $(s - \bar{s})(\mu_0^2) = 0$. Both effects are very 
small. The reader is referred to ref.~\cite{Catani:2004nc} for a further
discussion of especially the strange-sea asymmetry. 
%
%
\subsection{The treatment of heavy flavours}
\label{sec-hflav}
%
%
The evolution of the strong coupling and the parton distributions can
be performed in both the fixed flavour-number scheme (FFNS) and the 
variable flavour-number scheme (VFNS). This choice is made via the
initialization parameter {\tt IVFNS}. The former scheme is used for
{\tt IVFNS} = 0, any other value leads to the latter. The number of
flavours $\nf$ for the FFNS evolution is specified by the initialization
parameter {\tt NFF}. The values {\tt NFF} = $3 \ldots 6$ can be used. 
For {\tt IVFNS} $\neq$ 0 the number assigned to {\tt NFF} is irrelevant.

In the VFNS case we need the matching conditions between the effective
theories with $\nf$ and $\nf+1$ light flavours for both the strong 
coupling $\as$ and the parton distributions. For $\as$ these conditions
have been determined at N$^2$LO and N$^3$LO in refs.~\cite{Larin:1994va}
and \cite{Chetyrkin:1997sg}, for the unpolarized parton densities they 
are known to N$^2$LO from ref.~\cite{Buza:1996wv}.

In the present program the transitions $\nf \ra \nf+1$ are made, for
both $\as$ and the parton densities, when the factorization scale 
equals the pole masses of the heavy quarks,
\bea
\label{mu-nf1}
  \mu^2 \: = \: m_{\rm h}^2 \:\: , \qquad h \: = \: c, \: b, \: t 
  \:\: .
\eea
For this choice the matching conditions for the parton distributions
read up to N$^{m=2}$LO
\beq
\label{lp-nf1}
  l_i^{\,(\nf+1)}(x,m_h^2) \: = \:  l_i^{\,(\nf)}(x,m_h^2) +
  \delta_{m2} \: \as^2\: A^{\rm ns,(2)}_{qq,h}(x) \otimes
  l_i^{\, (\nf)}(x,m_h^2)
\eeq
where $l = q,\, \bar{q}$ and $i = 1,\ldots \nf$, and
\bea
\label{hp-nf1}
  g^{(\nf+1)}(x,m_h^2) \:\:\: &\! = \!\! &
    g^{\, (\nf)}(x,m_h^2) + \\[0.5mm] & &
    \delta_{m2} \, \as^2\, \Big[
    A_{\rm gq,h}^{S,(2)}(x) \otimes q_{\:\!\rm s}^{(\nf)}(x,m_h^2) +
    A_{\rm gg,h}^{S,(2)}(x) \otimes g^{(\nf)}(x,m_h^2) \Big ]
  \nn \\[0.5mm]
  (h+\bar{h})^{(\nf+1)}(x,m_h^2) \! &\! =\!\! &
    \delta_{m2} \, \as^2\: \Big [
    \tilde{A}_{\rm hq}^{S,(2)}(x)\otimes q_{\:\!\rm s}^{(\nf)}(x,m_h^2) 
    + \tilde{A}_{\rm hg}^{S,(2)}(x)\otimes g^{(\nf)}(x,m_h^2) \Big  ]
  \quad \nonumber
\end{eqnarray}
with $h=\bar{h}$. The coefficients $A^{(2)}$ for the spin-averaged case
can be found in Appendix B of ref.~\cite{Buza:1996wv} from where also 
their notation has been taken over. Due to our choice (\ref{mu-nf1}) 
for the thresholds only the scale-independent parts of the expressions
for $A^{(2)}$ are needed. 

The corresponding N$^{\rm m\,}$LO relation for the coupling constant 
at $\mu_r^2 = \kappa \mu^2$ is given by
\bea
\label{as-nf1}
  \as^{\, (\nf+1)}(\kappa m_h^2) \: = \:
  \as^{\, (\nf)} (\kappa m_h^2) + \sum_{n=1}^{m} \,
  \Big( \as^{\, (\nf)}(\kappa m_h^2) \Big)^{n+1}
  \sum_{l=0}^{n} \, c^{}_{n,l} \, \ln^{\, l} \kappa \:\: .
\eea
The pole-mass coefficients $c_{n\leq3,l}$ in Eq.~(\ref{as-nf1}) can 
be inferred from Eq.~(9) of Ref.~\cite{Chetyrkin:1997sg}, where $4\, 
\as^{\, (\nf-1)}$ is expressed in terms of $4\,\as^{\, (\nf)}$,
i.e., the matching is written backward in $\nf$ for a different 
normalization of the coupling. In our notation these coefficients
read
\bea
\label{as-cf1}
  c^{}_{1,0} & = & \:\: 0 \: \:\:, \quad 
  c^{}_{1,1} \: = \: \: \frac{2}{3} \nn \\[1mm]
  c^{}_{2,0} & = & \frac{14}{3} \:\:, \quad
  c^{}_{2,1} \: = \: \frac{38}{3} \:\:, \quad
  c^{}_{2,2} \: = \: \frac{4}{9}
\eea
and
\bea
\label{as-cf2}
  c^{}_{3,0} & = & 340.729 - 16.7981\,\nf  \nn \\[1mm]
  c^{}_{3,1} & = & \frac{8941}{27} - \frac{409}{27}\,\nf \:\: , \quad
  c^{}_{3,2} \: = \: \frac{511}{9} \:\:, \quad
  c^{}_{3,3} \: = \: \frac{8}{27} \:\: .
\eea
As in Eq.~(\ref{beta-exp}), we have truncated the irrational N$^3$LO 
coefficients to six digits here. 
Note that, while the parton distributions are continuous at the flavour
thresholds for our choice
(\ref{mu-nf1}) up to NLO, the same holds for the NLO coupling constant
only under the additional condition $\mu_r = \mu$ leaving only the 
vanishing coefficient $c^{}_{1,0}$ in Eq.~(\ref{as-nf1}).
At NNLO we use $\as^{\,(\nf+1)}(\kappa m_h^2)$ on the right-hand sides 
of Eqs.~(\ref{lp-nf1}) and (\ref{hp-nf1}).

If the program is run in the variable flavour-number mode, the initial 
distributions are specified for $\nf=3$, and Eq.~(\ref{hp-nf1}) is 
employed at least for the charm distributions. Therefore the initial 
factorization scale $\mu_0$ has to satisfy $\mu_0 \leq m_c$ and, 
consequently, not too small a value should be chosen for $\kappa$. The
inclusion of top (and bottom) among the partons can be switched off by
simply assigning a sufficiently large value to the respective mass. 
The masses $m_h$, like $\mu_0$, the initial coupling $\as(\mu_0^2)$ and
the initial light-parton distributions, are not fixed at the 
initialization of the program, but are defined (and can, of course, be 
re-defined) at a later stage as explained below.
%
%
\setlength{\baselineskip}{0.54cm}
\section{Complex moments and the Mellin inversion}
\setcounter{equation}{0}
In this section we discuss the issues related to the inverse Mellin 
transformation required for recovering the $x$-space parton 
distributions (and, in general, related observables) from 
the $N$-space expressions used for the intermediate calculations. 
The discussion includes our choice of the Mellin-inversion contour, the
required analytic continuations and the restrictions on the form of the
initial distributions resulting from our approach.
\subsection{From moments to \boldmath{$x$}-space}
\label{sec-minv}
\begin{figure}[htbp] \unitlength 1mm
\begin{center}
\vspace*{-2mm}
\begin{picture}(160,70)
\put(20,-5){\epsfig{file=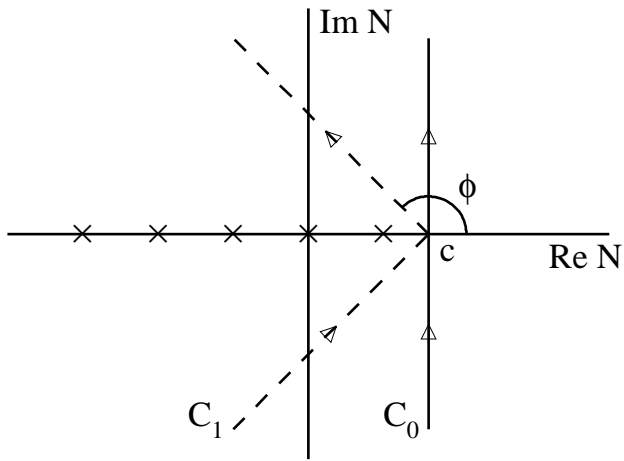,width=10cm}}
\end{picture}
\end{center}
\vspace*{-1mm}
\caption{Two integration contours for the inverse Mellin transformation
 in Eq.~(\protect\ref{M-inv}). The crosses schematically indicate the 
 singularities of the moment-space function $a(N)$.}
\vspace*{2mm}
\label{avfig1}
\end{figure}

%
%
We first consider the inverse transformation of the Mellin moments 
(\ref{N-trf}). If, as in our cases, $a(x)$ is piecewise smooth for 
$x>0$, the corresponding Mellin inversion reads
\beq
\label{M-inv}
  a(x) \: = \: \frac{1}{2\pi i} \int_{c-i\infty}^{c+i\infty} \! dN \: 
               x^{-N} a(N) \:\: ,
\eeq
where the real number $c$ has to be such that $ \int_{0}^{1} dx \, 
x^{\,c-1} a(x)$ is absolutely convergent \cite{CourHilb}. Hence $c$ has 
to lie to the right of the rightmost singularity $N_{\rm max}$ of 
$a(N)$. The contour of the integration in Eq.~(\ref{M-inv}) is displayed
in Fig.~\ref{avfig1} and denoted by ${\cal C}_{0}$. Also shown is a 
deformed route ${\cal C}_{1}$, yielding the same result as long as no 
singularities $N_i$ of $a(N)$ are enclosed by ${\cal C}_{0} - 
{\cal C}_{1}$. The $N_i$ are real with $ N_i < N_{\rm max} < c $ for 
the \NmLO\ evolution of parton distributions, thus this requirement is 
fulfilled automatically in our case.

It is useful to rewrite Eq.~(\ref{M-inv}) as an integration over a real
variable. We are dealing with functions which satisfy $ a^{\ast}(N) = 
a(N^{\ast})$, where $\ast$ denotes the complex conjugation. 
In this case contours characterized by the abscissa $c$ and the angle 
$\phi $ as in Fig.~\ref{avfig1} yield
\beq
\label{m-inv1}
  xa(x) \: = \: \frac{1}{\pi} \int_{0}^{\infty} \! dz\:\mbox{Im} \left[ 
  \, e^{i\phi} \, x^{1-c-z\exp{(i\phi)}} \: a(N=c+ze^{i\phi})\, \right] 
  \:\: .
\eeq
While formally the integral does not depend on $c > N_{\rm max}$ and 
$\pi > \phi \geq \pi/2$, a suitable choice of these parameters is
obviously useful for an efficient numerical evaluation. In particular,
it is advantageous to choose $ \phi > \pi/2 $ since then the factor 
$\exp \left(z \log{\frac{1}{x}} \cos{\phi}\right)$ introduces an 
exponential dampening of the integrand with increasing $z$, in addition 
to the oscillations already present for $\phi = \pi/2$, i.e., the
textbook contour ${\cal C}_{0}$. Consequently a smaller upper limit 
$z_{\rm max}$ can be employed in the numerical implementation of 
Eq.~(\ref{m-inv1}). 

The contour implemented in the present program forms a practitioner's
compromise between speed and accuracy for the cases encountered in 
QCD evolutions. The same contour is employed for all inversions, and
on that contour a fixed set of intervals is defined. For each of these
intervals we perform a standard eight-point Gauss-Legendre integration
\cite{AbrSteg1}.
As said, all this is done regardless of $x$, the initial distributions 
and all evolution parameters discussed in the previous section. Hence
all moments of the splitting functions, evolution-operator 
coefficients and matching conditions need to be calculated only once
on the resulting fixed grid of complex \mbox{$N$-values} at the 
initialization of the program. Note that this approach is rather 
different from that of refs.~\cite{Kosower:1997hg,Weinzierl:2002mv}, 
where a parabolic contour is optimized with respect to the shape of
the initial distributions.

Specifically, we choose $\phi = 3/4\: \pi$ (too close to neither  
${\cal C}_{0}$ nor the singularities of the integrands) and $c = 1.9$
(see below). The upper limit $z_{\rm max}$ depends on $x$ in accordance
with the $\ln (1/x)$ damping mentioned below Eq.~(\ref{m-inv1}) above: 
Eight intervals covering the region $0< z< 5$ are used for $x< 0.01$
(of which four have $0< z< 1$). 
Three more intervals with $5< z< 14$ are added for $0.01\leq x < 0.3$,
and another three covering $14< z< 32$ are included for $0.3\leq x < 
0.7$.  Above the latter value of $x$ the final four intervals with 
$32< z< 80$ are included as well. Under these conditions the chosen 
value $c = 1.9$ for the abscissa is a compromise between high accuracy 
for very small quantities at very large $x$ (improved for larger 
values) and very small $x$ (improved for smaller values).
 
This standard setup, used for the default initialization parameter {\tt
IFAST} = 0 is sufficient for a five-digit accuracy of the evolution
of the proton structure at $10^{-7} \leq x \leq 0.9$, with the 
exception of the tiny antiquark distributions at $x \simeq 0.9$. A
yet faster, but at very large and very small $x$-values less reliable
inversion can be invoked by {\tt IFAST} $\neq$ 0. The program then runs
with minimally 4 and maximally 10 (instead of 8 and 18) intervals. 
%
%
\subsection{Splitting functions and matching coefficients}
\label{sec-PtoN}
%
%
The complex-$N$ moments required for the evolution cannot be
computed by direct numerical integrations of Eq.~(\ref{N-trf}) on
our tilted contour ${\cal C}_{1}$ shown in Fig.~\ref{avfig1}.
Therefore we have to work with the proper analytic continuations 
or, where these are not known, with sufficiently accurate $x$-space
parametrizations of which the moments are known analytically. That
situation already occurs at NLO, but becomes more severe at higher 
orders.

The unpolarized NLO splitting functions have been programmed, long ago,
using the \mbox{integer-$N$} results in Eq.~(2.5) of ref.~\cite
{Gonzalez-Arroyo:1979df} (see also section 5 of ref.~\cite
{Curci:1980uw}) and in Appendix B of ref.~\cite{Floratos:1981hs}, 
together with the analytic continuations provided in Appendix A.1 of 
ref.~\cite{Gluck:1989ze}. 
With one exception, these analytic continuations can be expressed in 
terms of the complex polygamma functions $\psi^{(n)}(z)$. These 
functions are calculated from their asymptotic expansions 
\cite{AbrSteg2}, after for $|{\rm Im}(z)| < 10$ using the functional 
equations to shift the argument to ${\rm Re} (z) > 10$. The exception 
are the moments
\beq
\label{li-mom}
   \int_0^1 \!dx \: x^{N-1} \frac{Li_2(x)}{1+x} 
\eeq
where $Li_2$ is the standard dilogarithm. Improving on Eq.~(A.6) of 
ref.~\cite{Gluck:1989ze}, these moments can be approximated to a 
sufficient accuracy by inserting the parametrization
\bea
\label{Li2-par}
   \frac{1}{1+x} 
   & \!\cong\! & 1 - 0.9992\, x + 0.9851\, x^2 - 0.9005\, x^3 
   \nn \\[-0.5mm] & & \: \mbox{} 
                 + 0.6621\, x^4 - 0.3174\, x^5 + 0.0699\, x^6 
\eea
into Eq.~(\ref{li-mom}) and using the standard expression 
\cite{Devoto:1983tc} for the moments of $Li_2(x)$. A  useful list of 
Mellin transforms can also be found in the appendix of 
ref.~\cite{Blumlein:1998if}, see also ref.~\cite{Blumlein:2000hw}.
The polarized NLO splitting functions 
\cite{Mertig:1996ny,Vogelsang:1996vh,Vogelsang:1996im}
involve the same set of functions; they have been implemented using 
the appendix of ref.~\cite{Gluck:1995yr}. 

The recent complete integer-$N$ expressions for the unpolarized NNLO 
splitting functions \cite{Moch:2004pa,Vogt:2004mw} in moment space 
include harmonic sums \cite{Vermaseren:1998uu} up to weight five, of 
which the analytic continuations are not known at present. We therefore 
resort to the accurate $x$-space parametrizations of these functions 
provided in Eqs.~(4.22) -- (4.24) of ref.~\cite{Moch:2004pa} and 
Eqs.~(4.32) -- (4.35) of ref.~\cite{Vogt:2004mw}. The complex moments 
of these approximations can be expressed in terms of the polygamma 
functions. Based on the accuracy of these parametrizations and the size 
of the NNLO effects in the evolution~\cite{Moch:2004pa,Vogt:2004mw}, we 
expect that the relative errors introduced by this procedure amount to 
about $10^{-4}$ or less.

Finally we need, for the NNLO variable flavour-number evolution, the 
complex moments of the flavour-matching coefficients $A^{(2)}$ of 
ref.~\cite{Buza:1996wv}. These can be expressed in terms of the
psi-functions $\psi^{(n)}(z)$ as well, with the exception of $\tilde{A}
_{\rm hg}^{S,(2)}(x)$ in Eq.~(\ref{hp-nf1}). Also for this function 
the program presently uses the moments of a parametrization, viz
\bea
\label{a2-par}
   \tilde{A}_{\rm hg}^{S,(2)}(x) & \!\cong\! & \mbox{} 
   - 1.111\: \ln^3(1-x) - 0.400\: \ln^2(1-x) - 2.770\: \ln(1-x)
   \nn \\ & & \mbox{}
   - 187.8 + 249.6\: x - 146.8\: \ln^2 x \ln(1-x) - 93.68\: \ln x
   \nn \\ & & \mbox{}
   - 3.292\: \ln^2 x - 1.556\: \ln^3 x - 24.89\: x^{-1} 
   - 0.006\: \delta (1-x) \:\: . \:\: 
\eea
The convolution of the approximation (\ref{a2-par}) with typical gluon
distributions differs from the exact results by less than one part in
a thousand except close to zeros of these results. Note that the 
parametrization (\ref{a2-par}) has been employed for the approximate 
\cite{vanNeerven:2000wp} VFNS NNLO part of the 2001/2 evolution 
benchmarks presented in table 6 of ref.~\cite{lh2001a}. 
%
%
\subsection{The initial distributions}
%
%
Also the initial conditions $f_i(x,\mu_0^2)$ for the evolution are, of 
course, needed by the program in a form which facilitates a computation 
of the moments on the Mellin inversion contour. 
This is, presumably, the most severe restriction of the flexibility of 
the $N$-space approach as compared to direct numerical $x$-space 
solutions. For the latter, all quantities are usually defined on 
discrete grids of $x$-values, hence no functional form whatsoever is 
required for $f_i(x,\mu_0^2)$. For the present program a functional 
form is definitely needed, and furthermore that form should preferably 
be such that its complex moments can be readily computed.  This is, for 
example, not the case for the form employed in the CTEQ$\,$6 analysis
\cite{CTEQ6}, 
\beq
\label{CTEQ6}
  xf_i(x,\mu_0^2) \:\ = \: N_i\, x^{a_i} (1-x)^{b_i}\, 
  \left( 1 + A_i\, x \right)^{c_i} \, e^{\:\!d_ix} \:\: .
\eeq
An $N$-space evolution of Eq.~(\ref{CTEQ6}) would require an accurate
internal re-parametrization.

What can be readily handled by an $N$-space program, on the other
hand, should be perfectly adequate for a sufficiently general ansatz
for the initial distributions. In the present version of the program,
we have included the two six-parameter standard forms
\beq
\label{input1}
  xf_i(x,\mu_0^2) \: =\: N_i\: p_{i,1}^{}\: x^{\:\!p_{i,2}^{}} 
  (1-x)^{p_{i,3}^{}}\, \left[\, 1 + p_{i,5}^{}\, x^{\:\!p_{i,4}^{}} 
  + p_{i,6}^{}\, x \, \right] 
\eeq
and
\beq
\label{input2}
  xf_i(x,\mu_0^2) \: =\: N_i\: p_{i,1}^{}\: x^{\:\!p_{i,2}^{}} 
  (1-x)^{p_{i,3}^{}}\, \left[\, 1 + p_{i,4}^{}\, x^{0.5} 
  + p_{i,5}^{}\, x + p_{i,6}^{}\, x^{1.5}\, \right] \:\: .
\eeq
The moments of these functions are given in terms of Euler's Beta 
function $B(z_1,z_2)$. This functions is implemented using the 
asymptotic expansion of the logarithm of the Gamma function (the Stirling
formula) \cite{AbrSteg2} at $ {\rm Re} (z_1) > 5 $ and $ {\rm Re} (z_2) > 5 $ together 
with the functional equation. The $N$-dependent argument $z_1$ is first
inverted for $ {\rm Re} (z_1) < -10 $, i.e., the asymptotic expansion 
is invoked for $z_1^{\,\prime} = 1 - z_1 - z_2$. 
 
Speed is a much more important issue here than for the initialization 
of the splitting functions, $U$-matrix elements and matching conditions:
In fits of the parton densities, we need to deal efficiently with a 
large number of calls of the initial distributions, each of which in 
turn requires on the order of $10^3$ evaluations of $B(z_1,z_2)$. 
Note that the last term in the square bracket in Eq.~(\ref{input1}),
like the corresponding last two terms in Eq.~(\ref{input2}), does not
require new calls of $B(z_1,z_2)$ since the functional equation in the
first argument can be used instead. Therefore the extension of Eqs.\
(\ref{input1}) and (\ref{input2}) to (many) suitably chosen higher
powers of $x$ does not pose any efficiency problems. 

The ansatz (\ref{input1}) or (\ref{input2}) is used for the initial 
$u$ and $d$ valence-quark distributions $u_{\rm v}=u-\bar{u}$, 
$d_{\rm v}=d-\bar{d}$, the corresponding antiquark (sea) densities
$L_+ = 2(\bar{u}+\bar{d}\,)$ and $L_- = \bar{d} - \bar{u}$, the gluon
distribution $g$, and for the strange-flavour combinations $s_{\pm} = 
s \pm \bar{s}$. 
Heavy-flavour initial distributions $h(x,\mu_0^2)$ are obviously not 
required for the VFNS evolution starting with $\nf = 3$, see section 
\ref{sec-hflav}.  At present, independent shapes for $h(x,\mu_0^2)$ are
not included for the FFNS evolution either. The definitions and 
available user-options for the extra coefficients $N_i$ in 
Eqs.~(\ref{input1}) and (\ref{input2}) will be explained below.
%
%
\section{A brief {\sc Qcd-Pegasus} user guide}
\setcounter{equation}{0}
Most of the routines building up the evolution package are not directly
called by the user for standard applications. All he/she needs to 
interact with, are the routines for the initialization of the program,
the specification of the initial distributions and parameters like
$\as(\mu_0^2)$, and for reading out the results of the evolution.
In this section we describe the available input and output variables of 
these routines and show a small example program.
\subsection{General initialization}
\label{sec-init}
%
%
All input-independent quantities required for the unpolarized evolution 
(the polarized case is deferred to section 4.5) and $\nf$-matching 
described the section 2 are initialized~by

{\tt CALL INITEVOL(`EVOLPAR')}  .

\nin
The integer parameter {\tt EVOLPAR} defines how values are assigned to 
the six initialization parameters {\tt NPORD}, {\tt FR2} introduced in 
section \ref{sec-pevol}, {\tt IMODEV} discussed in sections 
\ref{sec-nevol} and \ref{sec-nsing}, {\tt IVFNS} and {\tt NFF} of 
section \ref{sec-hflav} and {\tt IFAST} defined in section 
\ref{sec-minv}. For {\tt EVOLPAR} = 1, {\tt INITEVOL} reads these 
parameters from a six-line file {\tt { usrinit.dat }} looking like
\vspace{-2.5mm}
\begin{verbatim}
  0        IFAST
  1        IVFNS
  4        NFF
  1        IMODEV
  1        NPORD
  1.0D0    FR2  .
\end{verbatim}
\vspace{-2.5mm}
For {\tt EVOLPAR} = 2, {\tt INITEVOL} obtains the corresponding values 
by calling the subroutine

{\tt USRINIT (IFAST, IVFNS, NFF, IMODEV, NPORD, FR2)} 

\nin
provided by the file {\tt { usrinit.f }}. For any other other value 
of {\tt EVOLPAR}, the program uses a default set of initialization 
parameters, actually the values displayed above. 

The available initialization options can be briefly summarized as 
follows$\,$:

\nin
{\tt IFAST} \\[1mm]
Speed/accuracy flag for the Mellin inversion. {\tt IFAST} = 0 is the 
standard. Any other value leads to a faster, but especially for very 
large and very small $x$ less reliable version. 

\nin
{\tt IVFNS} \\[1mm]
Switches between the evolution with a fixed number of flavours
(for {\tt IVFNS} = 0) and that in the variable flavour-number scheme,
starting with $\nf = 3$ (for any other value). 

\nin
{\tt NFF} \\[1mm]
The fixed number $\nf$ of flavours (between 3 and 5) for {\tt IVFNS} 
= 0.  Unused for {\tt IVFNS} $\neq$ 0.

\nin
{\tt IMODEV} \\[1mm]
Switches between the various evolution modes beyond LO. {\tt IMODEV} = 
1 and 2 invoke the respective iterated solutions of the evolution 
equations truncated in $\partial / \partial \ln \mu^2$ and $\partial / 
\partial \as$. Other values ({\tt IMODEV} = 3 is special) lead to the 
faster truncated solutions. {\tt IMODEV} = 1 emulates the standard 
$x$-space treatment if $\as(\mu_r^2)$ is adequate, i.e., not truncated
itself.

\nin
{\tt NPORD} \\[1mm]
The perturbative order $m = 0, 1$ or 2 of the evolution, defined as the
`m' in  \NmLO.

\nin
{\tt FR2} \\[1mm]
The constant ratio $\mu^2/\mu_r^2$ of the factorization scale $\mu$ and 
the renormalization scale $\mu_r$.
%
%
\subsection{Input parameters and initial distributions}
\label{sec-inp}
%
%
The input parameters and initial light-parton distributions for the 
evolution are set by

{\tt CALL INITINP(`INPPAR')} .

\nin
Analogous to {\tt EVOLPAR} in the previous section, the integer 
parameter {\tt INPPAR} specifies how the input parameters are read in.
If {\tt INPPAR} is neither 1 nor 2, the program will evolve a default
input, viz the one used for the 2001/2 benchmark tables~\cite{lh2001a}, 
\bea
\label{std-inp}
  xu_{\rm v}(x,\mu_0^2) &\! =\! & 5.107200\: x^{0.8}\: (1-x)^3 \nn\\
  xd_{\rm v}(x,\mu_0^2) &\! =\! & 3.064320\: x^{0.8}\: (1-x)^4 \nn\\
  xg\,(x,\mu_0^2)       &\! =\! & 1.700000\, x^{-0.1} (1-x)^5  \\
  x\bar{d}\,(x,\mu_0^2) &\! =\! & .1939875\, x^{-0.1} (1-x)^6  \nn\\
  x\bar{u}\,(x,\mu_0^2) &\! =\! & (1-x)\: x\bar{d}\,(x,\mu_0^2)
  \nn\\
  xs\,(x,\mu_0^2)       &\! =\! & x\bar{s}\,(x,\mu_0^2)
    \: = \: 0.2\, x(\bar{u}+\bar{d}\,)(x,\mu_0^2)
    \nn
\eea
with 
\beq
\label{std-as}
  \alpha_{\rm s}(\mu_{0}^2\! =\! 2\mbox{ GeV}^2) \: = \: 0.35 
\eeq
and, in the variable flavour-number case 
\bea
\label{std-hqm}
  m_{\rm c} \: = \: \mu_{0}          \: , \quad
  m_{\rm b} \: = \: 4.5 \mbox{ GeV}  \: , \quad
  m_{\rm t} \: = \: 175 \mbox{ GeV}  \:\: .
\eea
{\tt INITINP} reads these parameters from a file {\tt { usrinp.dat }}
for {\tt INPPAR} = 1. For the above input (using the ansatz 
(\ref{input1}) for definiteness) that file may look like
\vspace{-2mm}
\begin{verbatim}
  2.0D0       M20
  0.35D0      ALPHSI
  2.0D0       MC2
  20.25D0     MB2
  3.0625D4    MT2
  1           NFORM
  1           IMOMIN
  0           ISSIMP
  2.0D0,       0.8D0,    3.0D0,    0.5D0,    0.0D0,    0.0D0      PUV
  1.0D0,       0.8D0,    4.0D0,    0.5D0,    0.0D0,    0.0D0      PDV
  0.193987D0,  0.9D0,    6.0D0,    0.5D0,    0.0D0,    0.0D0      PLM
  0.136565D0, -0.1D0,    6.0D0,    0.5D0,    0.0D0,   -0.5D0      PLP
  0.0D0,       0.9D0,    6.0D0,    0.5D0,    0.0D0,    0.0D0      PSM
  0.027313D0, -0.1D0,    6.0D0,    0.5D0,    0.0D0,   -0.5D0      PSP
  1.0D0,      -0.1D0,    5.0D0,    0.5D0,    0.0D0,    0.0D0      PGL .
\end{verbatim}
\vspace{-2mm}
For {\tt INPPAR} = 2, {\tt INITINP} obtains the corresponding values
by calling the subroutine
\vspace{-2mm}
\begin{verbatim}
       USRINP (PUV, PDV, PDL, PDL, PLS, PSS, PGL, M20, ALPHSI, 
     ,         MC2, MB2, MT2, NFORM, IMOMIN, ISSIMP)
\end{verbatim}
\vspace{-2mm}

\nin
provided, for example, by the file {\tt { usrinp.f }}. This subroutine 
can also be modified to form a fit-parameter interface to programs like 
{\sc Minuit}.

The input parameters and options for the initial distributions 
presently available are$\,$:

\nin
{\tt M20} \\[1mm]
The initial factorization scale $\mu_0^2$. Equal to or smaller than 
$m_c^2$ for the VFNS~evolution.

\nin
{\tt ALPHSI} \\[1mm]
The strong coupling $\als$ [{\sc not} our internal $\as \equiv 
\als/(4\pi)$] at {\tt M20} = $\mu_0^2$ (even for $\mu_r \neq \mu$).

\nin
{\tt MC2 < MB2 < MT2}\\[1mm]
The squared {\it c, b} and {\it t} masses for the VFNS case 
{\tt IVFNS} $\neq$ 0. Irrelevant for {\tt IVFNS} = 0.

\nin
{\tt NFORM} \\[1mm]
Switches between the functional forms (\ref{input1}) for {\tt NFORM} 
= 1, and (\ref{input2}) for {\tt NFORM} = 2.

\nin
{\tt IMOMIN} \\[1mm]
For {\tt IMOMIN} = 0, $N_{L_{+}}$ = $N_{s_{+}}$ = 1 is used in Eqs.\
(\ref{input1}) and (\ref{input2}). Otherwise these factors are such
that $p^{}_{L_{+},1} =$ {\tt PLP(1)} and $p^{}_{{\rm s}_{+},1} =$ 
{\tt PSP(1)} are the corresponding momentum fractions.

\nin
{\tt ISSIMP} \\[1mm]
Switches between the full input ansatz (for {\tt ISSIMP} = 0) and a 
simplified input (otherwise), $s+\bar{s} =$ {\tt PSP(1)}$\cdot L_+$ 
with $s = \bar{s}$, for the initial strange-flavour distributions. 

\nin
{\tt PUV, PDV, PLM, PLP, PSM, PSP, PGL} \\[1mm]
The dimension-6 arrays for the input parameters $p^{}_{i,j}$ in 
Eqs.~(\ref{input1}) and (\ref{input2}) for $u_{\rm v}$, $d_{\rm v}$,
$L_-=\bar{d}-\bar{u}$, $L_+=2(\bar{d}+\bar{u})$, $s_\pm=s\pm\bar{s}$ 
and $g$. {\tt PUV(1)} and {\tt PDV(1)} are the
respective quark numbers (first moments), {\tt 2.D0} and {\tt 1.D0} 
for the case of the proton. {\tt PGL(1)} is the total fractional 
momentum carried by the partons (usually {\tt1.D0}). $N_{L_{-}}$ = 
$N_{\rm s_{-}}$ = 1. {\tt PSM(5)} is not used, the corresponding input 
parameter is fixed by the vanishing of the first moment. 
%
%
\subsection{The \boldmath{$x$}-space parton distributions}
\label{sec-inv}
%
%
The evolved parton distributions Mellin-inverted to $x$-space are 
accessed by

{\tt CALL XPARTON (PDFX, AS, `X', `M2', `IFLOW', `IFHIGH', `IPSTD')} .

\nin
Here {\tt PDFX} is the double-precision parton output array, 
user-declared before as {\tt PDFX(-6:6)}.$\!$ {\tt AS} (also 
double-precision, as all real variables) returns the corresponding 
value $\as(\mu_r^2(\mu^2))$. 
{\tt X} and {\tt M2} are simply the $x$ and $\mu^2$ for which
the evolved partons are requested from {\tt XPARTON}. The other three
(integer) parameters are flags helping to avoid wasting time.
For {\tt IPSTD} = 0, the notation for the index of {\tt PDFX}~is

{\tt PDFX(0)}  = $g$, \quad {\tt PDF(1)} = $u_{\rm v}$, \quad 
{\tt PDFX(-1)} = $u + \bar{u}$, \quad {\tt PDF(2)} = $d_{\rm v}$, 
\quad $\ldots$
 
\nin
and otherwise

{\tt PDFX(0)}  = $g$, \quad {\tt PDF(1)} = $u$, \quad$\,$ 
{\tt PDFX(-1)} = $\bar{u}$, \qquad$\:\,$ {\tt PDF(2)} = $d$, 
\quad $\ldots \:\: .$

\nin
Note that {\tt XPARTON} return $xg$ etc.
{\tt IFLOW} and {\tt IFHIGH} are lower and upper limits for the values
of this index for which the Mellin-inversion is actually performed.
For example, if the program is run for $\nf = 3$ parton flavours, there
is no point (except once, for checking) in letting the program
work to produce the inevitable zeros for $c(x,\mu^2)$ etc, thus 
{\tt IFLOW} = $-3$ and {\tt IFHIGH} = 3 should be used. If, moreover,
the user is interested (this time) only in the combinations $q_i + 
\bar{q}_i$ and the gluon density, then {\tt IPSTD} = 0, {\tt IFLOW} = 
$-3$ and {\tt IFHIGH} = 0 would be most efficient. Also the calculation
of the valence distributions is, in terms of speed, better performed 
using {\tt IPSTD} = 0.
%
%
\subsection{An example program with output}
\label{sec-ex}
%
%
We now present a small main program (provided by the file {\tt 
{ lh01tab.f }}) which can be used to check whether the evolution package
is working properly at least up to NLO. The internal standard inputs 
are used in the calls of both {\tt INITEVOL} and {\tt INITINP}. 
\begin{verbatim}
       PROGRAM LH01TAB
*
       IMPLICIT DOUBLE PRECISION (A - Z)
       DIMENSION PDFX(-6:6), XB(11)
       PARAMETER ( PI = 3.1415 92653 58979 D0 )
       INTEGER K1
*
* ..Access the input parton momentum fractions and normalizations
       COMMON / PANORM / AUV, ADV, ALS, ASS, AGL,
     ,                   NUV, NDV, NLS, NSS, NGL
*
* ..The values for mu^2 and x
       DATA M2 / 1.D4 /
       DATA XB / 1.D-7,  1.D-6,  1.D-5,  1.D-4,  1.D-3,  1.D-2,
     ,           1.D-1,  3.D-1,  5.D-1,  7.D-1,  9.D-1 /
*
* ..General initialization (internal default)
       CALL INITEVOL (0)
*
* ..Input initialization (internal default)
       CALL INITINP (0)
*
* ..Output of the momentum fractions and normalizations
  10   FORMAT (2X,'AUV =',F9.6,2X,'ADV =',F9.6,2X,'ALS =',F9.6,
     ,         2X,'ASS =',F9.6,2X,'AGL =',F9.6)
*
       WRITE(6,11) NUV, NDV, NLS, NSS, NGL
  11   FORMAT (2X,'NUV =',F9.6,2X,'NDV =',F9.6,2X,'NLS =',F9.6,
     ,         2X,'NSS =',F9.6,2X,'NGL =',F9.6,/)
*
* ..Loop only over x (M2 is fixed in the Les-Houches tables)
       DO 1 K1 = 1, 11
         X = XB (K1)
*
* ..Call of the Mellin inversion, reconstruction of L+ and L-
         CALL XPARTON (PDFX, AS, X, M2, -5, 2, 0)
         LMI = (PDFX(-2) - PDFX(-1) - PDFX(2) + PDFX(1)) * 0.5
         LPL =  PDFX(-1) + PDFX(-2) - PDFX(1) - PDFX(2)
*
* ..Output to be compared to the upper part of Table 4 
         IF (K1 .EQ. 1) WRITE (6,12) AS * 4*PI
  12     FORMAT (2X,'ALPHA_S(MR2(M2)) =  ',F9.6,/)
*
         IF (K1 .EQ. 1) WRITE (6,13)
  13     FORMAT (2X,'x',8X,'xu_v',7X,'xd_v',7X,'xL_-',7X,'xL_+',
     ,           7X,'xs_+',7X,'xc_+',7X,'xb_+',7X,'xg',/)
*
         WRITE (6,14) X, PDFX(1), PDFX(2), LMI, LPL, PDFX(-3),
     ,                   PDFX(-4), PDFX(-5), PDFX(0)
  14     FORMAT (1PE6.0,1X,8(1PE11.4))
*
   1   CONTINUE
       STOP
       END
\end{verbatim}
 
Here is what this program returns. The main table has been slightly 
edited (mainly {\tt E-07} $\ra$ {\tt E-7} etc.), to avoid having to use 
a yet smaller font.
 
\vspace*{-2.5mm}
\footnotesize
\begin{verbatim}
 AUV = 0.333333  ADV = 0.137931  ALS = 0.136565  ASS = 0.027313  AGL = 0.364858
 NUV = 5.107200  NDV = 3.064320  NLS = 0.775950  NSS = 0.155190  NGL = 1.700000

  ALPHA_S(MR2(M2)) =   0.116032

  x       xu_v      xd_v      xL_-      xL_+      xs_+      xc_+      xb_+      xg

1.E-7  1.0927E-4 6.4125E-5 4.3925E-6 1.3787E+2 6.7857E+1 6.7139E+1 6.0071E+1 1.1167E+3
1.E-6  5.5533E-4 3.2498E-4 1.9829E-5 6.9157E+1 3.3723E+1 3.3153E+1 2.8860E+1 5.2289E+2
1.E-5  2.7419E-3 1.5989E-3 8.5701E-5 3.2996E+1 1.5819E+1 1.5367E+1 1.2892E+1 2.2753E+2
1.E-4  1.3039E-2 7.5664E-3 3.5582E-4 1.4822E+1 6.8739E+0 6.5156E+0 5.1969E+0 8.9513E+1
1.E-3  5.8507E-2 3.3652E-2 1.4329E-3 6.1772E+0 2.6726E+0 2.3949E+0 1.7801E+0 3.0245E+1
1.E-2  2.3128E-1 1.2978E-1 5.3472E-3 2.2500E+0 8.4161E-1 6.5235E-1 4.3894E-1 7.7491E+0
1.E-1  5.5324E-1 2.7252E-1 9.9709E-3 3.9099E-1 1.1425E-1 6.0071E-2 3.5441E-2 8.5586E-1
3.E-1  3.5129E-1 1.3046E-1 3.0061E-3 3.5463E-2 9.1084E-3 3.3595E-3 1.9039E-3 7.9625E-2
5.E-1  1.2130E-1 3.1564E-2 3.7719E-4 2.3775E-3 5.7606E-4 1.6761E-4 1.0021E-4 7.7265E-3
7.E-1  2.0102E-2 3.0932E-3 1.3440E-5 5.2605E-5 1.2166E-5 2.7408E-6 2.0095E-6 3.7574E-4
9.E-1  3.5232E-4 1.7855E-5 8.6806E-9 2.0302E-8 3.9024E-9-2.638E-10 5.839E-10 1.1955E-6
\end{verbatim}
\normalsize
\vspace*{-2.5mm}

\setlength{\parskip}{0.3cm}
\setlength{\baselineskip}{0.55cm}
 
\nin
The first line of the output provides the respective momentum fractions 
carried by the valence quarks, the light-quark sea and the gluons at 
the initial scale. This information has been obtained by accessing the 
common-block {\tt PANORM} filled for this purpose by the inner input 
routine not described in this section. The second line are the 
corresponding normalization factors, cf.~Eq.~(\ref{std-inp}). 
The rest is a set of reference results which agree, except for some
marginal offsets at $x = 0.9$ for the tiny sea-quark distributions, 
with the upper part of table 4 in ref.\ \cite{lh2001a}. A test run 
by the user should lead to the same numbers.

A final remark on the efficient calculation of the parton distributions
at several scales. {\tt XPARTON} saves the moments used for the 
inversion, and will re-use them the next time --- recall that the $a(N)$
in Eq.~(\ref{m-inv1}) do not know about $x$ --- unless the input or the 
scale $\mu^2$ have been changed. Consequently one should first perform
the Mellin inversion for all desired $x$-values at one scale before
proceeding to the next scale, instead of ordering the calls of 
{\tt XPARTON} in another manner. The same applies, of course, also 
to corresponding routines determining observables like structure 
functions from the $N$-space expressions. 
%
%
\subsection{The longitudinally polarized case}
\label{sec-pol}
%
%
The evolution of the polarized parton distributions is performed  
completely analogous to the unpolarized case discussed above. Indeed, 
one could write the program such that the same user-interface routines
would deal with both cases. Here we have decided to keep at least the 
polarized and unpolarized initialization and input routines separately. 

{\tt CALL INITPOL(`EVOLPAR')} 

\nin
initialises the polarizated evolution. The options are almost identical
to those discussed in section \ref{sec-init}. The order of the evolution
is restricted to {\tt NPORD} = 0 and 1 at present, since the polarized 
three-loop splitting functions are not yet known. The respective files 
supplying the initialization parameter for {\tt EVOLPAR} = 1 and 2 are 
{\tt { usrpinit.dat }} and {\tt { usrpinit.f }} providing the subroutine
{\tt USRPINIT}. The defaults are the same as in section~\ref{sec-init}.

The input parameters and initial polarized distributions for the
evolution are set by

{\tt CALL INITPINP(`INPPAR')} .

\nin
The corresponding data files are {\tt { usrpinp.dat }} and 
{\tt { usrpinp.f }} containing the routine {\tt USRPINP}. Instead of
Eq.~(\ref{std-inp}) the default toy input for program checks reads
\bea
\label{pol-inp}
  x\Delta u_{\rm v} &\! =\! & +1.3\: x^{0.7}\: (1-x)^3\: (1+3x) \nn\\
  x\Delta d_{\rm v} &\! =\! & -0.5\: x^{0.7}\: (1-x)^4\: (1+4x) \nn \\
  x\Delta g\:       &\! =\! & +1.5\: x^{0.5}\: (1-x)^5  \nn \\
  x\Delta\bar{d}\:  &\! =\! & x\Delta\bar{u} 
     \: = \: -0.05\: x^{0.3}\: (1-x)^7  \nn\\
  x\Delta s\:       &\! =\! & x\Delta\bar{s} 
     \: = \: +0.5\: x\Delta \bar{d} \:\: .
\eea
The other input parameter have to same meaning (and the same defaults)
as in section \ref{sec-inp}, with the exception of 

\nin
{\tt IMOMIN}\\[1mm]
For {\tt IMOMIN} = 0 (used as the internal default here), we put $N_i = 
1$ in Eqs.~(\ref{input1}) and (\ref{input2}) for all seven input 
combinations $u_{\rm v}$, $d_{\rm v}$, $L_-=\bar{d}-\bar{u}$, 
$L_+=2(\bar{d}+\bar{u})$, $s_\pm=s\pm\bar{s}$ and $g$. Otherwise all 
$N_i$ are chosen such that the first elements of {\tt PUV}, {\tt PDV}, 
{\tt PLM}, {\tt PLP}, {\tt PSM}, {\tt PSP} and {\tt PGL} represent the 
first moments of the respective initial distributions. {\tt PSM(5)} is
a normal input parameter in the polarized case.
%
%
\subsection{Output in \boldmath{$N$}-space}
\label{sec-mom}
%
%
The above setup assumes that the user is interested in the $x$-space
parton distributions as obtained by the method outlined in section
\ref{sec-minv}. For those who want to evolve fixed moments (e.g., 
momentum fractions) or prefer to invert the $N$-space results by other 
means (e.g., Monte-Carlo integration), we also provide the 
user-interface subroutine

{\tt NPARTON (PDFN, AS, `N', `M2', `IPSTD', `IPOL', `EVOLPAR', 
              `INPPAR') .}

\nin
There is no point to have separate overall and input initialization
calls here, as all quantities need to be recalculated each time anyway 
as {\tt N} is a (double-complex) free parameter. Consequently the calls 
of the initialization and input routines have been integrated into 
{\tt NPARTON}, which therefore takes over the respective parameters 
{\tt EVOLPAR} and {\tt INPPAR} (the options remain as before). In this 
case also the choice between the unpolarized ({\tt IPOL} =0) and 
polarized evolution (otherwise) is made via the call of {\tt NPARTON}.
Note that this mode of using the program is an option for which the 
code has not really been optimized. 
The double-complex output array has to be declared as {\tt 
PDFN(-6:6)} in the calling program.
%
%
\setlength{\baselineskip}{0.53cm}
\section{A short reference guide}
In this section we provide brief descriptions of the subroutines and
functions included in the evolution package. Unless specified otherwise,
the function or subroutine `{\tt ROUTINE}' is stored in the file
`{\tt routine.f}', where more information can be found. This results in 
quite a few files, but it appears easier this way to keep track of 
future upgrades and user modifications.
A very large part of the internal communications in the program proceeds
via named common-blocks. A corresponding list is included at the end of 
this section. 
\subsection{Initialization routines}
\label{ref-init}
%
%
The main initialization routine for the $N$-space based parton 
evolution is

\vspace{-3mm}
\subsubsection{\tt INITEVOL(`IPAR')} 
\vspace{-3mm}
 
The options for {\tt IPAR} have already been discussed in section 
\ref{sec-init}. This routine sets some constants like the Euler-%
Mascheroni constant $\gamma_{\rm e} \equiv$ {\tt ZETA(1)}, the lowest 
values of the Riemann's Zeta function, $\zeta_{\,i>1} =$ {\tt ZETA(i)}, 
and the lowest SU(3) invariants {\tt CA}, {\tt CF} and {\tt TR}. 
It provides the fixed array {\tt NA} of complex moments and the 
corresponding weights {\tt WN} for the Gauss integrations discussed in
section \ref{sec-minv}. 
The analytically continued simple harmonic sums $S_i(N)$ are calculated
for $i=1,\ldots ,6$ on these support points and stored as {\tt S(K,i)}
where {\tt K} is the parameter of the array {\tt NA}. Finally the 
routine calls, depending on the initialization parameters discussed in 
section \ref{sec-init}, the subroutines for the $N$-space splitting 
functions, $\nf$-matching conditions and evolution matrices listed 
below.

{\tt INITEVOL} also sets a couple of internal initialization parameters.
The consistency of the order {\tt NAORD} of the coupling constant (see 
section \ref{sec-as}) with that of the parton evolution is enforced by 
{\tt NAORD = NPORD}. The number of steps for the Runge-Kutta integration
of Eq.~(\ref{as-eqn}) and the maximal power of $\as$ in the $U$-matrix 
solution (\ref{u-sol}) are set to

{\tt NASTPS = 20 } and { \tt NUORD = 15} . 

\nin
Note that {\tt NUORD} is presently restricted by array declarations to 
values of 20 or less. 

\vspace{-2.5mm}
\subsubsection{\tt USRINIT (IFAST, IVFNS, NFF, IMODEV, NPORD, FR2)}
\vspace{-3mm}

The subroutine providing the initialization parameters if {\tt 
INITEVOL} is called with {\tt IPAR}$\,= 2$.

\vspace{-2.5mm}
\subsubsection{\tt BETAFCT}
\vspace{-3mm}
 
Provides the values {\tt BETA(0)}, $\ldots$, {\tt BETA(3)} 
(\ref{beta-exp}) of the QCD $\beta$-function for $\nf = 3, \ldots, 6$. 

\vspace{-2mm}
\subsubsection{\tt PSI(Z), DPSI(Z,M)\quad in the file\quad 
               psifcts.f}
\vspace{-3mm}
 
The complex $\psi$-function {\tt PSI(Z)} and its $m$-th derivatives 
$\psi^{(m)}(z) =$ {\tt DPSI(Z,M)} calculated using the functional
equations and the asymptotic expansions as discussed in section~\ref
{sec-PtoN}.

\setlength{\baselineskip}{0.55cm}
\vspace{-2mm}
\subsubsection{\tt PNS0MOM, PSG0MOM\quad in the file\quad 
               pnsg0mom.f}
\vspace{-2mm}
  
The lowest-order non-singlet (routine {\tt PNS0MOM}) and singlet
(routine {\tt PSG0MOM}) \mbox{$N$-space} splitting function {\tt P0NS} 
and {\tt P0SG} on the array {\tt NA} with parameter {\tt KN} defined by 
{\tt INITEVOL}, as all splitting functions for $\nf = 3, \ldots, 6$. 
The singlet sector uses (also at higher orders) the matrix notation 
(\ref{ev-sg}) with the array arguments {\tt 1 = q} and {\tt 2 = g}.

\vspace{-2mm}
\subsubsection{\tt PNS1MOM} 
\vspace{-2mm}
  
The subroutine for the NLO non-singlet splitting functions {\tt P1NS}
at $\mu_r = \mu$ on the array {\tt NA}.
The routine needs to be called before {\tt PSG1MOM}, since it provides 
part of $P^{\,(1)}_{\rm qq}$, see Eq.~(\ref{Pqq}), and the non-trivial 
analytic continuations according to Appendix (A.1) of 
ref.~\cite{Gluck:1989ze} and Eq.~(\ref{Li2-par}). Starting from NLO the 
three cases (\ref{P-ns}) are included via an additional array dimension  
with arguments {\tt 1 = +}, {\tt 2 = -} and {\tt 3 = v} ({\tt = -} at 
NLO).

\vspace{-2mm}
\subsubsection{\tt PSG1MOM} 
\vspace{-2mm}
  
The corresponding subroutine for the NLO singlet splitting functions 
{\tt P1SG(KN,NF,i,j)}.

\vspace{-2mm}
\subsubsection{\tt PNS2MOM}
\vspace{-2mm}

The NNLO non-singlet splitting functions {\tt P2NS} in $N$-space, as 
always on the array {\tt NA}, as obtained from the accurate $x$-space 
parametrizations (4.22)--(4.24) of ref.~\cite{Moch:2004pa} since the 
complex moments of the exact results are presently unknown. Analogous
to the NLO case, this routine needs to be called before the singlet
case {\tt PSG2MOM}. Also here $\mu_r = \mu$.

\vspace{-2mm}
\subsubsection{\tt PSG2MOM}
\vspace{-2mm}

The corresponding routine for the singlet quantities {\tt P2SG}, using
(4.32)--(4.35) of ref.~\cite{Vogt:2004mw}. 

\vspace{-2mm}
\subsubsection{\tt LSGMOM}
\vspace{-2mm}

The eigenvalue decomposition (\ref{lo-dec1})--(\ref{lo-dec3}) of the LO 
singlet splitting-function matrix {\tt P0SG} divided by {\tt BETA0}.
The $N$- and $\nf$-dependent eigenvalues are denoted by 
{\tt R(KN,NF,l)}, the corresponding projection operators by 
{\tt E(KN,NF,i,j,l)} with {\tt l = 1, 2}.

\vspace{-2mm}
\subsubsection{\tt USG1MOM}
\vspace{-2mm}

The subroutine for the NLO singlet evolution matrix $\UV_{\!1} $= 
{\tt U1(KN,NF,I,J)} in $N$-space obtained from Eq.~(\ref{U-res}) 
for~$k=1$.  Here and in the following routines the terms arising from 
$\mu_r \neq \mu$ in Eq.~(\ref{P-exp2}) are included.
This routine needs to be called before {\tt USG1HMOM}.
Procedures (presently absent) for scheme transformations of the 
splitting functions (provided in \MSbar\ by the above routines) would 
have to be called before this routine.

\vspace{-2mm}
\subsubsection{\tt USG1HMOM}
\vspace{-2mm}

Adds the higher-order contributions $\UV_{\!k\,} $= {\tt U1H(K,..)} to the 
$U$-matrices for the iterated NLO solutions of the evolution equations
discussed in section \ref{sec-nevol}. Note that there are no 
non-singlet routines corresponding to {\tt USG1MOM} and {\tt USG1HMOM}
since $U_1^{\rm \pm}$ are trivial and Eqs.~(\ref{ns-sol1}) and 
(\ref{ns-sol2}) are used for the iterated non-singlet solutions instead 
of Eq.~(\ref{u-sol}).

\vspace{-2mm}
\subsubsection{\tt UNS2MOM}
\vspace{-2mm}

The routine for the NNLO non-singlet evolution operators {\tt UNS2},
including the higher-order pieces for the iterative solutions. Also 
here the quantities for the evolution of $q_{\rm ns}^{\,\pm{\rm v}}$
are included by an extra array dimension with the argument
{\tt 1 = +}, {\tt 2 = -} and {\tt 3 = v}. 

\vspace{-2mm}
\subsubsection{\tt USG2MOM}
\vspace{-2mm}

As the subroutine {\tt USG1MOM}, but providing {\tt U2} for the 
truncated NNLO solution ($k=2$).

\vspace{-2mm}
\subsubsection{\tt USG2HMOM}
\vspace{-2mm}

As {\tt USG1HMOM}, but adding the higher terms {\tt U2H} required for 
the iterated NNLO solutions. The singlet $U$-matrix routines partly 
build upon each other, hence they should be called in a proper order, 
like the one used in this list.  
 
\vspace{-2mm}
\subsubsection{\tt ANS2MOM}
\vspace{-2mm}

The $\as^{\:\!2}$ (NNLO) non-singlet coefficient 
$A^{\rm ns,(2)}_{qq,h} =$ {\tt A2NS(KN)} for the \MSbar\ flavour-number 
transition (\ref{lp-nf1}) at $\mu^2=m_h^2$, obtained from the 
$x$-space results in Appendix B of ref.~\cite{Buza:1996wv}.

\vspace{-2mm}
\subsubsection{\tt ASG2MOM}
\vspace{-2mm}

The $\as^{\:\!2}$ (NNLO) $N$-space singlet coefficients {\tt A2SG} in 
Eq.~(\ref{hp-nf1}) obtained from the same source. The routine uses 
{\tt A2NS}, hence it is to be called after {\tt ANS2MOM}. 
The moments of the parametrization (\ref{a2-par}) are used for 
$\tilde{A}_{\rm hg}^{S,(2)} $= {\tt A2SG(KN,1,2)}.
%
%
\subsection{Input and flavour-threshold routines}
\label{ref-inp}
%
%
The steering routine for the $N$-space distributions at initial scale 
and flavour thresholds~is

\vspace{-2mm}
\subsubsection{\tt INITINP(`IPAR')}
\vspace{-2mm}

The options switched by {\tt IPAR} have already been discussed in 
section \ref{sec-inp}. This routine calls the appropriate routine
{\tt INPLMOM1} or {\tt INPLMOM2} and calculates the initial value
of $\as$ for the evolution, which is unequal to the (properly 
normalized) input parameter in section~\ref{sec-inp} for $\mu_r\neq\mu$.
For {\tt IVFNS} $\neq 0$, {\tt EVNFTHR} is then used to store also the 
parton distributions at the heavy-flavour thresholds $\mu^2 = m_h^2$. 
This is another efficiency measure: Recall that the evolution is 
performed in fixed-$\nf$ steps, using Eqs.~(\ref{lp-nf1})--%
(\ref{as-nf1}) in between. For all values $\mu^2 > m_{b\,}^2$, e.g., 
the evolution to $\mu^2 =m_b^2$ is thus the same for a given input, and 
there is no point to repeat this part of the computation for every new
value of $\mu^2$.

\vspace{-2mm}
\subsubsection{\tt USRINP (PUV,...,PGL, M20, ASI,
                          MC2,MB2,MT2, NFORM,IMOMIN,ISSIMP)}
\vspace{-2mm}

The subroutine providing the input parameters if {\tt INITINP} is 
called with {\tt IPAR}$\,= 2$.

\vspace{-2mm}
\subsubsection{\tt INPLMOM1 (PUV, PDV, PLM, PLP, PSM, PSP, PGL, 
 IMOMIN, ISSIMP)}
\vspace{-2mm} 

This routine calculates, from the input parameters discussed already in
section \ref{sec-inp}, the moments of the ansatz (\ref{input1}) on the
array {\tt NA} and stores them in the basis (\ref{ns-basis}) with the
notation $q_{\:\!\rm ns}^{\:\!\rm v}(N)$ = {\tt VAI(KN)}, 
$q_{\:\!\rm s}$ = {\tt SGI}, $v_k^-$ = {\tt M`k'I}, $v_k^+$ = 
{\tt P`k'I} and $g$ = GLI for $k=3, 8$. 
Also the common-block with the second moments and normalizations shown 
in section \ref{sec-ex} is written here. The routine sets the flag 
{\tt IINNEW = 1} to inform other routines, like {\tt XPARTON} discussed
in \ref{sec-inv}, that a new input call has taken place.

\vspace{-2mm}
\subsubsection{\tt INPLMOM2 (PUV, PDV, PLM, PLP, PSM, PSP, PGL, 
 IMOMIN, ISSIMP)}
\vspace{-2mm}

As the previous routine, but for the ansatz (\ref{input2}) for the 
three-flavour initial distributions. Other input forms can be added as 
{\tt INPLMOM3} etc.~if needed, which requires only a miminal additional
modification of {\tt INITINP} in section 5.2.1.

\vspace{-2mm}
\subsubsection{\tt EBETA(Z1,Z2)\quad in the file\quad ebetafct.f}
\vspace{-2mm}

The complex Beta function {\tt EBETA(Z1,Z2)} calculated from the 
asymptotic expansion of $\ln \Gamma(z)$ as discussed in section 
\ref{sec-PtoN}. Called (only) by the routines for the $N$-space
inputs.

\vspace{-2mm}
\subsubsection{\tt EVNFTHR (MC2, MB2, MT2)}
\vspace{-2mm}

For {\tt IVFNS} = 1, this routine is used to evolve $\as$ and the
partons from the three-flavour initial scale to the four- to 
six-flavour thresholds $\mu^2 = m_h^2$. The results for $\as$ are 
stored as {\tt ASC}, {\tt ASB} and {\tt AST}, those for $q_{\:\!\rm ns}
^{\:\!\rm v}$ as {\tt VAC}, {\tt VAB} and {\tt VAT}, etc.
For $m_t^2 > 10^{10}$ or $m_{b,t}^2 > 10^{10}$ (in GeV$^2$, as always),
the corresponding parts of the calculations are skipped.

\vspace{-2mm}
\subsubsection{\tt ASNF1 (ASNF, LOGRH, NF)\quad in the file\quad
 asmatch.f}
\vspace{-2mm}

This functions computes $\as^{\, (\nf+1)}$ = {\tt ASNF1} from 
$\as^{\, (\nf)}$ = {\tt ASNF} and $\ln (\mu_r^2/m_h^2)$ = {\tt LOGRH}
according to Eqs.~(\ref{as-nf1})--(\ref{as-cf1}) at maximally N$^3$LO.
The order of the $\nf$-matching is set by {\tt NAORD} specified via 
the initialization parameter {\tt NPORD}, see section 5.1.1. 
%
%
\setlength{\baselineskip}{0.53cm}
\subsection{Evolution and Mellin-inversion routines}
\label{ref-evol}
%
 
\subsubsection{\tt AS (R2, R20, AS0, NF)\quad in the file\quad asrgkt.f}
\vspace{-2mm}

Returns $\as$ = {\tt AS} at the renormalization scale $\mu_r^2$ = 
{\tt R2} as obtained beyond LO from a fourth order Runge-Kutta 
integration of Eq.~(\ref{as-eqn}) in {\tt NASTPS} steps from the initial
value $a_0$ = {\tt AS0} at $\mu_{r,0}^2$ = {\tt R20} at order {\tt 
NAORD} for {\tt NF} quark flavours. At LO Eq.~(\ref{as-impl}) is used.
Possible other versions have to be provided (in the same notation) by
the use her-/himself.

\vspace{-2mm}
\subsubsection{\tt ENSG0N (ENS, ESG, ASI, ASF, S, KN, NF)} 
\vspace{-2mm}

The subroutine for the LO non-singlet and singlet $N$-space evolution 
operators (\ref{lo-sol}) at a fixed number of flavours  {\tt NF}.
The respective kernels  {\tt ENS}  and {\tt ESG(I,J)}, {\tt I}, {\tt J} 
= 1, 2 with  {\tt 1 = q, 2 = g}, are returned for a moment $N$ 
specified via the counter {\tt KN} on the array {\tt NA}.  
The initial and final scales are specified by the respective values
$a_0$ = {\tt ASI} and $\as$ =  {\tt ASF} of the coupling constant.
{\tt S} = $\ln (a_0/\as)$ is not calculated internally for efficiency.

\vspace{-2mm}
\subsubsection{\tt ENS1N (ENS, ALPI, ALPF, S, KN, NF)}
\vspace{-2mm}

This routine returns one of the NLO hadronic non-singlet kernels {\tt 
ENS(K)} (\ref{ns-trnc})--(\ref{ns-sol3}) for the evolution of the `+' 
({\tt K=1}) and $-=\rm v$ ({\tt K=2,3}) combinations (\ref{q-ns}) of 
quark distributions. The mode of the evolution is set via the initialization
parameter {\tt IMODEV}. The input arguments of the routine are as 
described for the previous routine.

\vspace{-2mm}
\subsubsection{\tt ESG1N (ESG, ASI, ASF, S, KN, NF)}
\vspace{-2mm}

As {\tt ENS1N}, but for the singlet kernels {\tt ESG(I,J)} according to
Eqs.~(\ref{u-sol}) and (\ref{f-trnc}).

\vspace{-2mm}
\subsubsection{\tt ENS2N (ENS, ASI, ASF, S, KN, NF, NSMIN, NSMAX))}
\vspace{-2mm}

As {\tt ENS1N}, but for the NNLO kernels {\tt ENS(K)} using
Eqs.~(\ref{u-sol}), (\ref{f-trnc}) and (\ref{ns-trnc}). 
The additional arguments {\tt NSMIN} and {\tt NSMAX} set the range in
{\tt K} for which {\tt ENS(K)} is defined. This allows to save time, 
e.g., in e.m.~structure-function calculations requiring only {\tt ENS(1)}.

\vspace{-2mm}
\subsubsection{\tt ESG2N (ESG, ASI, ASF, S, KN, NF)}
\vspace{-2mm}

As {\tt ESG1N}, but for the NNLO kernels {\tt ESG(I,J)} governing the
evolution of the flavour-singlet parton distributions. Recall that
all these kernels assume a fixed ratio $\mu/\mu_r$.
 
\vspace{-2mm}
\subsubsection{\tt EVNVFN (PDFN, ASI, ASF, NF, NLOW, NHIGH, IPSTD)}
\vspace{-2mm}

The subroutine providing the $N$-space parton distributions {\tt PDFN}
for the variable flavour-number evolution. These results are returned
for the part {\tt NLOW < KN < NHIGH} of the array ${\tt NA}$ in a 
notation chosen via {\tt IPSTD} as discussed in section \ref{sec-inv}.
The routine makes use of the flavour-threshold results stored by 
{\tt EVNFTHR} described in section 5.2.6.

\vspace{-2mm}
\subsubsection{\tt EVNFFN (PDFN, ASI, ASF, NF, NLOW, NHIGH, IPSTD)}
\vspace{-2mm}

As {\tt EVNVFN}, but for the fixed flavour-number evolution with
$3 \leq$ {\tt NF} $\leq 5$ partonic flavours.

\vspace{-2mm}
\subsubsection{\tt XPARTON (PDFX, AS, `X', `M2', `IFLOW', `IFHIGH', 
 `IPSTD')} 
\vspace{-2mm}

The top-level evolution and Mellin-inversion routine for standard use
of the program, discussed already in section \ref{sec-inv}. It checks 
whether the previous results for $\as$ and {\tt PDFN} can be re-used 
(same {\tt M2} and no new input call), calculates the otherwise 
required quantities, calls {\tt EVNVFN} or {\tt EVNFFN} if necessary,
and carries out the Gauss integrations.
%
%
\subsection{Routines for the polarized case}
\label{ref-pol}
%

To avoid a substantial duplication of code at this point, the polarized
evolution uses the internal routines described above as far as possible.
Thus a simultaneous evolution of unpolarized and polarized 
parton distributions, e.g., for $N$-space analyses of $pp$ scattering 
with only one proton polarized, is not possible at this point. 
The additional routines are:

\vspace{-2mm}
\subsubsection{\tt INITPOL(`IPAR')}
\vspace{-3mm}

The polarized version of the main initialization routine {\tt INITEVOL}
in section 5.1.1. The same options are available except for the 
restriction of the evolution to LO and NLO.

\vspace{-2mm}
\subsubsection{\tt USRPINIT (IFAST, IVFNS, NFF, IMODEV, NPORD, FR2)}
\vspace{-2mm}

The subroutine providing the input parameters for {\tt INITPOL} if
called with {\tt IPAR}$\,= 2$.

\vspace{-2mm}
\subsubsection{\tt PSG0PMOM}
\vspace{-3mm}

The lowest-order polarized singlet \mbox{$N$-space} splitting function 
{\tt P0SG} corresponding to the unpolarized routine in section 5.1.5.
Note that there is no corresponding non-singlet routine, as in this
case the unpolarized and polarized splitting functions are identical.  

\vspace{-2mm}
\subsubsection{\tt PNS1PMOM, PSG1PMOM}
\vspace{-3mm}

The polarized counterparts of {\tt PNS1MOM} and {\tt PSG1MOM} described
in sections 5.1.6 and 5.1.7. The same array names and arguments are
employed for use by the general $U$-matrix and evolution routines in 
sections \ref{ref-init} and \ref{ref-evol}. The non-singlet quantities
are identical to the unpolarized case, but with the `+' and { \tt - = 
v } entries interchanged.

\vspace{-2mm}
\subsubsection{\tt ANS2PMOM, ASG2PMOM}
\vspace{-2mm}

The subroutines for the polarized case corresponding to {\tt ANS2MOM} 
and {\tt ASG2MOM} in sections 5.1.16 and 5.1.17. Presently these are 
dummy routines. The presence of the arrays {\tt A2NS} and {\tt A2SG} is 
technically required in {\tt EVNFTHR} (section 5.2.6). 

\vspace{-2mm}
\subsubsection{\tt INITPINP(`IPAR')}
\vspace{-2mm}

The input initialization routine for the polarized evolution 
corresponding to {\tt INITINP} in section 5.2.1. {\tt IPAR} switches the
input options as discussed in section \ref{sec-inp} and \ref{sec-pol}.

\vspace{-2mm}
\subsubsection{\tt USRPINP (PUV,...,PGL, M20, ASI,
                          MC2,MB2,MT2, NFORM,IMOMIN,ISSIMP)}
\vspace{-2mm}

The subroutine providing the input parameters for {\tt INITPINP} if
called with {\tt IPAR}$\,= 2$.

\vspace{-2mm}
\subsubsection{\tt INPPMOM1, INPPMOM2} 
\vspace{-2mm}

The input-moment subroutines corresponding to {\tt INPLMOM1} and 
{\tt INPLMOM2} in sections 5.2.3 and 5.2.4. The only difference of the 
input options has been discussed in section~\ref{sec-pol}. The
common-block {\tt PANORM} now returns the first instead of the 
second moments as {\tt AUV} etc. 
%
%
\subsection{Routines for output in \boldmath{$N$}-space}
%
%
The user interface for obtaining the results of the evolution at any
complex value of $N$ is

\vspace{-3mm}
\subsubsection{\tt NPARTON (PDFN, AS, `N', M2', `IPSTD', `IPOL', 
                            `EVOLPAR', `INPPAR')} 
\vspace{-2mm}

briefly discussed in section \ref{sec-mom}. {\tt PDFN(-6:6)} is the
double complex output array with two options switched by {\tt IPSTD} 
as explained for the analogous $x$-space array {\tt PDFX} in section 
\ref{sec-inv}. {\tt AS} returns the corresponding value 
$\as(\mu_r^2(\mu^2))$ with $\mu^2$ = {\tt M2} (both double-precision).
Note that {\tt NPARTON} is called without previous initialization and
input calls as these calls are issues by this routine using the values
of {\tt IPOL}, {\tt EVOLPAR} and {\tt INPPAR}.

\vspace{-3mm}
\subsubsection{\tt INITMOM (IPOL, EVOLPAR)}
\vspace{-2mm}

The special initialization routine called by {\tt NPARTON}. Depending 
on {\tt IPOL} this routine takes over the respective roles of {\tt 
INITEVOL} in section 6.1.1 or {\tt INITPOL} in section 6.4.1 in the 
calculations for one moment $N$. The array dimension {\tt NA} for the 
standard setup is formally kept in order to comply with the fixed 
declarations in the low-level routines.
%
%
\subsection{A list of common blocks}
%
%
Here we present the list of common blocks mentioned above, ordered 
alphabetically. The dimensions of the array variables are not 
indicated for brevity. 
Most, but not all of the variables stored by these common blocks have 
been mentioned above.
\vspace{-2mm}
\begin{verbatim}
      COMMON / ANS2   / A2NS 
      COMMON / ASFTHR / ASC, M2C, ASB, M2B, AST, M2T
      COMMON / ASG2   / A2SG 
      COMMON / ASINP  / AS0, M20
      COMMON / ASPAR  / NAORD, NASTPS
      COMMON / BETA   / BETA0, BETA1, BETA2, BETA3
      COMMON / COLOUR / CF, CA, TR
      COMMON / EVMOD  / IMODEV
      COMMON / FRRAT  / LOGFR
      COMMON / HSUMS  / S
      COMMON / INPNEW / IINNEW
      COMMON / INVFST / IFAST
      COMMON / ITORD  / NUORD
      COMMON / KRON2D / D
      COMMON / LSG    / R
      COMMON / MOMS   / NA  
      COMMON / NCONT  / C, CC
      COMMON / NFFIX  / NFF
      COMMON / NFUSED / NFLOW, NFHIGH
      COMMON / NNUSED / NMAX 
      COMMON / ORDER  / NPORD
      COMMON / PABTHR / VAB, M3B, M8B, M15B, M24B, SGB, P3B, P8B, P15B,
     ,                  P24B, GLB
      COMMON / PACTHR / VAC, M3C, M8C, M15C, SGC, P3C, P8C, P15C, GLC
      COMMON / PAINP  / VAI, M3I, M8I, SGI, P3I, P8I, GLI 
      COMMON / PANORM / AUV, ADV, ALS, ASS, AGL, NUV, NDV, NLS, NSS, NGL
      COMMON / PATTHR / VAT, M3T, M8T, M15T, M24T, M35T, SGT, P3T, P8T, 
     ,                  P15T, P24T, P35T, GLT
      COMMON / PNS0   / P0NS
      COMMON / PNS1   / P1NS
      COMMON / PNS2   / P2NS
      COMMON / PSG0   / P0SG
      COMMON / PSG1   / P1SG
      COMMON / PSG2   / P2SG
      COMMON / R1SG   / R1
      COMMON / R2SG   / R2
      COMMON / RZETA  / ZETA
      COMMON / SPSUMS / SSCHLP, SSTR2P, SSTR3P
      COMMON / U1HSG  / U1H
      COMMON / U1SG   / U1
      COMMON / U2HSG  / U2H
      COMMON / U2NS   / UNS2
      COMMON / U2SG   / U2
      COMMON / VARFLV / IVFNS 
      COMMON / WEIGHTS/ WN
\end{verbatim}
\vspace*{-2mm}
 
%
%
\section{Accuracy and speed of the program}
\label{sec-acc}
%
%
The main numerical step in an $N$-space evolution program is the 
inverse Mellin transformation of the final results back to $x$-space. 
As discussed in section \ref{sec-minv}, the present program uses a
one-fits-all grid of fixed $N$-values for optimal performance in large
computations. The accuracy
of the programmed inversion can be easiest tested by calling 
{\tt XPARTON} at this initial scale {\tt M2} $ = \mu_0^2$ and dividing 
by the known $x$-space initial distributions.
This has been done in Fig.~2 for the two extreme shapes in 
Eq.~(\ref{std-inp}), $u_{\rm v}$ and $u_{\rm s} \equiv \bar{u}$. Recall
that $xu_{\rm v}$ vanishes as $x^{0.8}$ for $x \ra 0$, while $xu_{\rm s}
\sim (1-x)^7$ for $x \ra 1$. 

\begin{figure}[htb]
\vspace{1mm}
\centerline{\epsfig{file=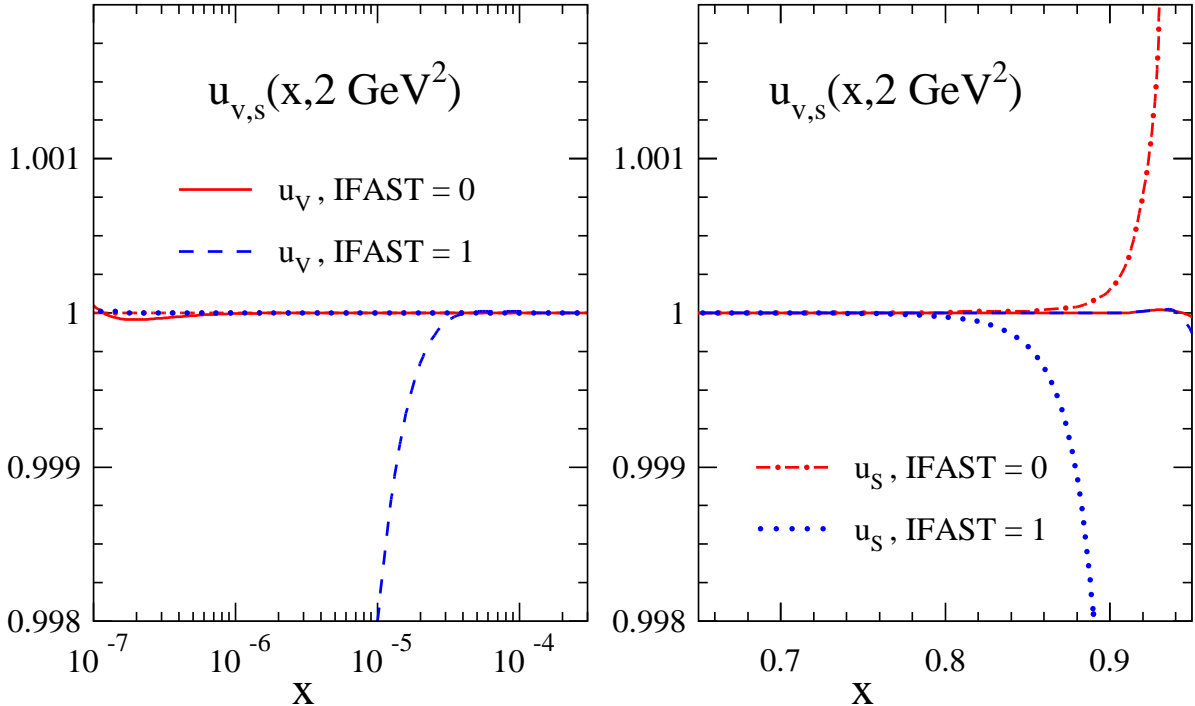,width=16cm,angle=0}}
\vspace{-3mm}
\caption{The up-valence and up-sea distributions at the initial scale
 (\ref{std-as}) obtained by the standard and fast Mellin inversions, 
 divided by the exact $x$-space inputs in Eq.~(\ref{std-inp}). The same
 four curves are shown in the left (right) part for small (large) 
 values of $x$.}
\vspace*{1mm}
\end{figure}

The results for the standard inversion using up to 144 moments 
({\tt IFAST} = 0) for this test do not show any noticeable deviations 
from the exact values at $10^{-6} < x < 0.9$ for $u_{\rm v}$, and at 
$10^{-11} < x < 0.8$ for $u_{\rm s}$. 
These ranges refer to the program's default value $c = 1.9$ of the 
contour abscissa in Eq.~(\ref{m-inv1}). Smaller values would improve
the inversion of $u_{\rm v}$ at extremely small $x$ at the cost of 
worsening that of $u_{\rm s}$ at very large $x$.
  
The faster, but close to the end-points inevitably less reliable option 
using between 32 and 80 moments ({\tt IFAST} = 1) can be safely employed
at least for $10^{-4}\,\lsim\, x\,\lsim\, 0.7$, where the upper limit 
includes the effect of the further softening of the sea-quark 
distributions by the evolution.
In many applications the small (and anyway poorly known) distributions
$xu_{\rm v}$ at smaller $x$ and $u_{\rm s}$ at larger $x$ are 
practically irrelevant; then the fast inversion can be used over the
full range of $x$ shown in the figure.

At NNLO the present overall accuracy of the program at `normal' values 
of $x$ is dominated by that of the parametrizations of the corresponding
splitting functions and the \mbox{$\nf$-matching} coefficient discussed
in section~\ref{sec-PtoN}. An update of the NNLO benchmark results in 
ref.~\cite{lh2001a} with the complete three-loop splitting functions
of refs.~\cite{Moch:2004pa,Vogt:2004mw} will be presented elsewhere 
\cite{SVprep}. For checks of tables 5 and 6 in ref.~\cite{lh2001a} the
user needs to employ the previous approximations \cite
{vanNeerven:2000wp} instead of the routines 5.1.8 and 5.1.9. For the 
moment these approximations are still available as { \tt pns2mold.f } 
and { \tt psg2mold.f }.

Finally we illustrate the speed of the program. For this purpose the 
input (\ref{std-inp}) is evolved, using Eqs.~(\ref{std-as}) and 
(\ref{std-hqm}), for variable $\nf$ to 20 scales $\mu^2$ and the results
are Mellin-inverted at 25 values of $x$ at each scale. The chosen scales 
and values of $x$ are
\vspace*{-1mm}
\begin{verbatim}
       DATA MS / 2.0D0,  2.7D0,  3.6D0,  5.D0,   7.D0,   1.D1,
     1           1.4D1,  2.D1,   3.D1,   5.D1,   7.D1,   1.D2,
     2           2.D2,   5.D2,   1.D3,   3.D3,   1.D4,   4.D4,
     3           2.D5,   1.D6 /
       DATA XB / 1.D-8,  1.D-7,  1.D-6,
     1           1.D-5,  2.D-5,  5.D-5,  1.D-4,  2.D-4,  5.D-4,
     1           1.D-3,  2.D-3,  5.D-3,  1.D-2,  2.D-2,  5.D-2,
     2           1.D-1, 1.5D-1,  2.D-1,  3.D-1,  4.D-1,  5.D-1,
     3           6.D-1,  7.D-1,  8.D-1,  9.D-1 / .
\end{verbatim}
\vspace*{-1mm}
The whole calculation, including the input initialization but excluding
the general initialization, is then repeated 200 times. Thus we perform
precisely $10^{\,5}$ calls of {\tt XPARTON}. The times (in seconds) 
required for this computation on a 2.0 GHz Pentium-IV processor are 
listed below for various values of {\tt NPORD}, {\tt IMODEV} and 
{\tt IFAST} (defined in section \ref{sec-init}). The value of the 
latter parameter is given after the name of the compiler. We have used
the GNU compiler {\tt g77} (version 3.2-36) with {\tt -O2} and the 
(non-commercial)
`Intel Fortran Compiler 8.0 for Linux' {\tt ifort} with {\tt -xW}
(optimization with {\tt -O2} is the default for {\tt ifort}).
\vspace*{-4mm}
\begin{center}
\begin{tabular}{l c c c c}\\
{Type of evolution\quad} &{\tt g77,0} &{\tt g77,1} &{\tt ifort,0} 
 &{\tt ifort,1}\\[1mm]
\hline & & & & \\[-3mm]
 \qquad\qquad\quad LO       &  7.8  &  4.3 &  2.5 &  1.4  \\[1.5mm]
 {\tt IMODEV=0}, NLO        &  8.2  &  4.5 &  2.7 &  1.5  \\[0.5mm]
 {\tt IMODEV=0}, NNLO       &  9.2  &  5.0 &  3.2 &  1.8  \\[1.5mm]
 {\tt IMODEV=1}, NLO        & 10.7  &  5.8 &  3.8 &  2.1  \\[0.5mm]
 {\tt IMODEV=1}, NNLO       & 11.3  &  6.2 &  4.4 &  2.4  \\[1mm]\hline
\end{tabular}
\end{center}
\vspace*{2mm}

Thus roughly $10^{\,4}$ to $10^{\,5}$ Mellin inversions of the complete
set of partons can be performed per second under the above conditions, 
depending on the initialization parameters and the available compiler. 
As expected, the evolution becomes somewhat slower with increasing 
order, and the iterative solutions (involving high orders of the 
evolution matrices $U$) are slower than the truncated evolution. 

%
%
\setlength{\baselineskip}{0.53cm}
\section{Summary and outlook}
%
%
We have presented the fast, flexible and accurate {\sc Fortran} package
QCD-{\sc Pegasus} \mbox{solving} the evolution equations for the unpolarized
and polarized parton distributions of hadrons in perturbative QCD.
The code is designed such that the program is relatively fastest when
this is most needed, viz in large repetitive computations like fits of 
the initial distributions or error analyses 
according to refs.~\cite{Giele:1998gw,Giele:2001mr,lh2001b}.
In standard applications, the user needs to interact with only very
few top-level routines which communicate the evolution parameters and
initial distributions to, and the results of the evolution from the
lower-level routines actually performing the underlying evolution in
Mellin-$N$ space.

The full potential of the $N$-space approach is realized once not only
the parton evolution, but also the convolutions with the partonic
cross sections (coefficient functions) are performed (as 
multiplications of the moments) before Mellin-inverting back to 
$x$-space. Doing this poses no problems if the coefficient functions
are known (to a sufficient accuracy) in a form facilitating the 
calculation of their complex-$N$ moments. Even if this is not the case
--- for example, if coefficient functions are only known via an 
$x$-space Monte-Carlo programs including experimental cuts --- there 
are efficient methods to proceed in $N$-space, see
ref.~\cite{Graudenz:1996sk} and especially refs.~\cite{Kosower:1998vj,%
Stratmann:2001pb}. Sample routines for the latter procedure will be
presented elsewhere for representative observables in $ep$ and $pp/
p\bar{p}$ scattering. The user can then implement the processes he or
she is interested in along these lines. QCD-{\sc Pegasus} can of 
course also be used efficiently within `normal' $x$-space analyses.

If additional routines are included to deal with the additional
`pointlike' solution, the program can also be used for the QCD 
evolution of the parton distributions of the photon. These routines
largely exist, but have not been included in this first public version
of the program. An even easier extension would be the evolution of
also `timelike' distributions (fragmentation functions) of effectively
massless partons. Considerably larger modifications would be required
to include other (hypothetical) light coloured particles like a light
gluino which would lead to a enlarged $3\times3$ flavour-singlet
sector. This extension was actually present in an earlier incarnation
of this program, see ref.~\cite{Ruckl:1994bh}, but has not been pursued
anymore since the light-gluino window is now generally considered
closed.

The code of QCD-{\sc Pegasus} can be obtained by downloading the source
of this article from the {\tt hep-ph} preprint archive. Relevant changes
of the code, if required, will be communicated by replacing the article
in this archive.
The program is distributed under the GNU public license, with the 
obvious additional requirement that scientific publications (including
preprints) which have been prepared using the present code, or a 
modified version of it, should include a reference to the present 
article. The author can be reached by electronic mail for bug reports, 
comments and questions. 
\subsection*{Acknowledgments}
\vspace*{-3mm}
I am grateful to M. Botje, S. Moch and W. Vogelsang for critically 
reading the text. This work was supported by the Dutch 
Foundation for Fundamental Research of Matter~(FOM).$\!$
 
%
\small
\setlength{\baselineskip}{0.5cm}

\end{document}